\documentclass[12pt]{article}

\global\arraycolsep=1pt

\usepackage{color,amsbsy,amssymb,latexsym,amsfonts, amsmath,cancel}
\usepackage{braket}
\usepackage{mathrsfs}
\usepackage{graphicx}

\numberwithin{equation}{section}
\newcommand{\bel}[1]{\begin{equation}\label{#1}}                     
\newcommand{\bal}[1]{\begin{eqnarray}\label{#1}}   
\newcommand{\be}{\begin{equation}}               
\newcommand{\ba}{\begin{eqnarray}}           
\newcommand{\ee}{\end{equation}}
\newcommand{\ea}{\end{eqnarray}}

\newcommand{\im}{\mathrm{i}}

\newcommand{\de}{\mathrm{d}}

\renewcommand{\thefootnote}{\fnsymbol{footnote}}
\newcommand{\abs}[1]{\left| #1 \right|}
\newcommand{\hs}{\hspace{10mm}}
\newcommand{\orddif}[2]{\frac{\de #1}{\de #2}}
\newcommand{\pardif}[2]{\frac{\partial #1}{\partial #2}}
\def\Res{\ensuremath\mathop\mathrm{Res}}

\newcommand{\bea}{\begin{equation}}
\newcommand{\eea}{\end{equation}}

\newcommand{\Average}[1]{\Bigl\langle\!\!\Bigr\langle #1 \Bigr\rangle\!\! \Bigr\rangle}
\begin{document}

%
%
\begin{titlepage}
\begin{flushright}
\normalsize
~~~~
OCU-PHYS 352\\
April, 2011
\end{flushright}

\vspace{15pt}

\begin{center}
{\LARGE  $\epsilon$-Corrected  Seiberg-Witten Prepotential   }\\
\vspace{5pt}
{\LARGE
Obtained From }\\
\vspace{5pt}
{\LARGE Half
Genus Expansion in $\beta$-Deformed Matrix Model } \\
\end{center}

\vspace{23pt}

\begin{center}
{ H. Itoyama$^{a, b} $\footnote{e-mail: itoyama@sci.osaka-cu.ac.jp}
  and 
 N. Yonezawa$^b$\footnote{e-mail: yonezawa@sci.osaka-cu.ac.jp} 
}\\
%
\vspace{18pt}
%

$^a$ \it Department of Mathematics and Physics, Graduate School of Science\\
Osaka City University\\
\vspace{5pt}

$^b$ \it Osaka City University Advanced Mathematical Institute (OCAMI)

\vspace{5pt}

3-3-138, Sugimoto, Sumiyoshi-ku, Osaka, 558-8585, Japan \\

\end{center}
%
\vspace{20pt}
\begin{center}
Abstract\\
\end{center}
%
We consider the half-genus expansion of the resolvent function
    in the $\beta$-deformed matrix model with three-Penner potential
    under the AGT conjecture
and the $0d-4d$ dictionary.
The partition function of the model, after the specification of the
paths, becomes the DF conformal block for fixed $c$ and provides the
Nekrasov partition
function expanded both in $g_s = \sqrt{-\epsilon_1 \epsilon_2}$ and in
$\epsilon = \epsilon_1+\epsilon_2$.
Exploiting the explicit expressions for the lower terms of the free
energy extracted
from the above expansion, we derive the first few
$\epsilon$ corrections to the Seiberg-Witten prepotential in terms of
the parameters
of $SU(2)$, $N_{f} =4$, ${\cal N}= 2$ supersymmetric gauge theory.   

\vfill

\setcounter{footnote}{0}
\renewcommand{\thefootnote}{\arabic{footnote}}

\end{titlepage}

\renewcommand{\thefootnote}{\arabic{footnote}}
\setcounter{footnote}{0}

\section{Introduction}

One promising way to study genus expansion of free energy in matrix models is
    to solve by iteration the finite $N$ Schwinger-Dyson equation
    in the presence of an infinite number of couplings to single trace functions
        \cite{ACKM:1993,Akemann:1996zr}\cite{David90,MM90,AM90,IM91}.
The partition function then acts as a generating function\cite{IM91, FKN1991,DVV1991} and free energy to a given order is 
    obtained from the one-point resolvent function by the inversion with respect to the coupling function.
Some of the recent references in this direction include \cite{E0407,CE0504,CE0604,C1009}.
In this paper, we consider this procedure in the context of the AGT conjecture \cite{AGT,Wyllard},
    namely, the 2d-4d conformal field theory connection where matrices act as a bridge\cite{DV,IMO}.

The Seiberg-Witten prepotential \cite{Seiberg:1994rs,SW9408} has been a central object
    in the study of low-energy effective action (LEEA) of ${\cal N}=2$ susy gauge theories.
This object and its extension are adequately characterized by the integrability properties \cite{Integrability}.
In recent years, its recognition as free energy of matrix model has been successful
    in some cases\cite{Dijkgraaf:2002s,whitham}.

In the above mentioned connection, 
    we have, on the one hand, the four-point conformal block -- 
    the representation theoretic (model independent)
    quantity of the Virasoro algebra --
    whose integral representation \cite{DF} is given by the $\beta$ deformed ensemble of the one-matrix model
    after the specification of the paths \cite{MMS0911,MMS1001,IO5,IOY}.
On the other side, we have so called Nekrasov partiton function \cite{Nekrasov0206}
    arising from the instanton sum in the case of $SU(2)$, $N_{f} =4$.
For some decoupling cases, see \cite{gai0908,MMM0909a,MMM0909b,AM0910,EM1,1003Yanagida,EM2}.
The equivalence of these two has been proven in some cases \cite{FatLit,HJS}
    and so far been mainly studied as a series expansion 
    in $q$ -- the cross ratio on the one side
    and the exponentiated holomorphic coupling on the other \cite{MMM0907,MM0908a,MM0908b}.
In particular, the $0d-4d$ dictionary of the parameters has been established
    in \cite{IO5,IOY,MMM1003,CDV1010,MMS1011a,MMS1011b,AFLT1012},
exploiting the properties of the Selberg integral and the Jack symmetric function
    \cite{sel,mac1,kad,kan,I}.
For more recent developments, see for instance
\cite{
    BT0909,KPW,MS1004,AY1004,NX1005,Tai1006,NX1006,KMS1007,MY1009,BMS1010,MMS1010,MMM1011,
    BMTY1011,MMS1012,M1101,BB1102,MMPS1103,Popolitov1001,FHT1003,AM0912,Giribet1001}.
Our study here will provide an alternative direction to pursue,
    which is currently somewhat underdeveloped, and may supply useful data
    to the renewed consideration of topological field theory \cite{AHNT1003,Taki1007}.

In the next section,
    we present the half-genus expansion of the resolvent
    in the case of $\beta$ ensemble with three Penner potential.  
We regard $\epsilon = \epsilon_1+\epsilon_2$ to be order $g_s$,
    which treats via $g_s = \sqrt{-\epsilon_1 \epsilon_2}$  the two deformation parameters
    in an equal footing.
This implies that $Q_E = \frac{\epsilon}{g_s}$ is order $1$ and
    that the central charge $c$ is kept fixed throughout our procedure.
In intermediate steps, the double expansion both in $\epsilon$ and in $g_s$ naturally appear. 
The point, albeit being minor,
    where our consideration differs from more generic discussion
    is that an $\epsilon$ correction is present already at the original potential according to the $0d-4d$ dictionary.
In section three, we present the formula for $F_{0,1}$ as well as that of $F_{0,2}$.
The representation in terms of 4d data is discussed.
In Appendix \ref{sec:Appendix_on_Riemann_surfaces},
    we review a few materials taken from theory of Riemann surfaces.
In Appendix \ref{sec:Appendix_On_F_01}, the proof of our formula in section \ref{sec:F_01} is given
    while the proof of our formula in section \ref{sec:F_02} is given in Appendix \ref{sec:Appendix_On_F_02}.
We include in Appendix \ref{sec:Appendix_On_Planar_limit} the complete expression for the resolvent
    at the planar level $w_{0,0}(z)$ as it appears not seen anywhere in the literature.

\section{Half genus expansion of resolvent and free energy}

The partition function of the model that we consider in this paper is
\begin{align}
\begin{split}
Z=  \mathop{\int}_{\mathcal{D}_L\times\mathcal{D}_R} 
        \prod_{I'=1}^N \de \lambda_{I'}
          \ &  
    \Delta(\lambda_{J})^{2b_E^2} \  \  
    \exp\left[
        b_E\sum_{I=1}^N W(\lambda_I;\alpha_i,\mathrm{g}_\ell)
    \right],\\
\Delta(\lambda_{J})&=\prod_{I<I'}  \abs{\lambda_I-\lambda_I'},\\
\mathcal{D}_L
&= [0,q]^{N_L},\\
\mathcal{D}_R
&= [1,\infty)^{N_R}.
\end{split}
\end{align}
Here 
\begin{align*}
W(z;\alpha_i,\mathrm{g}_\ell)
&=\alpha_1 \log z
    + \alpha_2 \log (z-q)
    + \alpha_3 \log (z-1)+
    \sum_{\ell=0}^\infty \mathrm{g}_\ell \  (z-z_{0})^\ell,
\end{align*}
is the three-Penner potential and the last term is included to generate the resolvent and we will set to vanish at the end of the operation. 
For the choice of the integration paths and its role, see \cite{IO5}.

Let us recall the finite $N$ loop equation (Virasoro constraints\cite{
        David90,MM90,AM90,IM91%
        }):
\begin{align}
\begin{split}
0&= \frac{1}{Z} \mathop{\int}_{\mathcal{D}_L\times\mathcal{D}_R} 
        \prod_{I'=1}^N \de \lambda_{I'} \  
        \sum_{I=1}^N\pardif{}{\lambda_I  }
    \left(
        \frac{1}{z-\lambda_ I}\Delta(\lambda_J)^{2b_E^2} \  \  
        \exp\left[
                b_E\sum_{\bar{I}=1}^N W(\lambda_{\bar{I}};\alpha_i,\mathrm{g}_\ell)
        \right]
    \right)\\
&=Q_E\orddif{}{z}\Average{
        b_E\sum_{I=1}^N \frac{1}{z-\lambda_I}
    }
    + \Average{
        b_E \sum_{I=1}^N \frac{1}{z-\lambda_I}
        b_E \sum_{J=1}^N\frac{1}{z-\lambda_J}
    }\\
&\hs +W'(z;\alpha_i,\mathrm{g}_\ell)\Average{
        b_E\sum_{I=1}^N \frac{1}{z-\lambda_I}
    }
    -\Average{
        b_E \sum_{I=1}^N \frac{
                W'(z;\alpha_i,\mathrm{g}_\ell)-W'(\lambda_I;\alpha_i,\mathrm{g}_\ell)
        }{z-\lambda_I}
    },
\end{split}
\label{eq:loop_equation_0}
\end{align}
where $Q_E=b_E-1/b_E$.
We have denoted by $\Average{\ \cdot\ }$ an average with respect to $Z$.

In \cite{IO5}, the precise dictionary
        between the six matrix-model parameters at finite $N$
        and the six parameters of Nekrasov partition function
                of $N_f =4$, $SU(2)$, $\mathcal{N}=2$ case, namely,
                the 0d-4d counterpart of the AGT\cite{AGT} relation has been established.
It is 
\begin{align}
\begin{split}
\alpha_1&=\frac{m_2-m_1+ \epsilon}{g_s},\\
\alpha_2&=\frac{m_2+m_1}{g_s},\\
\alpha_3&=\frac{m_3+m_4}{g_s},\\
\alpha_4&=\frac{m_3-m_4+\epsilon}{g_s},
\end{split}
\ \ 
\begin{split}
b_E N_L&=-\frac{m_2-a}{g_s},\\
b_E N_R&=-\frac{m_3+a}{g_s},\\
b_E N\equiv b_E (N_L+N_R)&=-\frac{m_2+m_3}{g_s},\\
&
\end{split}
\ \ 
\begin{split}
b_E&=\frac{\epsilon_1}{g_s},\\
\frac{1}{b_E}&=-\frac{\epsilon_2}{g_s},\\
Q_E=b_E-\frac{1}{b_E}&=\frac{\epsilon_1+\epsilon_2}{g_s}=\frac{\epsilon}{g_s},\\
\varrho \equiv N g_s b_E&=-(m_2+m_3).
\end{split}
\label{eq:1d-4d_dictionary} 
\end{align}
In the gauge theory side, $g_s=\sqrt{-\epsilon_1\epsilon_2}$
        while in the matrix model side,
        $g_s$ appears as an auxiliary parameter
        in generating the large $N$ expansion.
We will first generate the double expansion of the one point resolvent
        both in $g_s$ and in $\epsilon$ and subsequently set $\epsilon = Q_E g_s$.
Throughout our procedure,
        $Q_E$ and hence the central charge $c=1-6Q_E^2$ will be kept finite.

Let us rescale the parameters as
\begin{align}
\begin{split}
g_s \alpha_i &\equiv \tilde{\alpha}_i \equiv \bar{\alpha}_i
    +\epsilon \left(\delta_{i,1}+\delta_{i,4}\right),\\
g_s \mathrm{g}_\ell &\equiv \tilde{\mathrm{g}}_\ell,
\end{split}\\
\begin{split}
W(z;\alpha_i,\mathrm{g}_\ell) 
&= \frac{1}{g_s}W(z;
                        \tilde{\alpha}_i=\bar{\alpha}_i+\epsilon \delta_{i,1},
                        \tilde{\mathrm{g}}_\ell
                )\\
&= \frac{1}{g_s}\Big(
                W_{0,0}(z)+\epsilon W_{0,1}(z) + J(z)
        \Big),
\end{split}
\end{align}
where
\begin{align*}
\begin{split}
W_{0,0}(z)&=\bar{\alpha}_1\log z +\bar{\alpha}_2 \log(z-q)+\bar{\alpha}_3\log(z-1),\\
W_{0,1}(z)&=\log(z),\\
J(z)&\equiv \sum_{\ell=0}^\infty \tilde{\mathrm{g}}_\ell (z-z_0)^\ell.
\end{split}
\end{align*}
Introduce the resolvent
\begin{align}
\hat{w}(z)\equiv g_s b_E \sum_{i=1}^N \frac{1}{z-\lambda_I}.
\end{align}
Multiplying (\ref{eq:loop_equation_0}) by $g_s^2$, we obtain
\begin{align}
\begin{split}
0&= \epsilon \orddif{}{z}\Average{
        \hat{w}(z)
    }
    + \Average{
        \hat{w}(z)^2
    }
    +W'(z;\tilde{\alpha}_i,\tilde{\mathrm{g}}_\ell )\Average{
        \hat{w}(z)
    }
    -\Average{
        \hat{f}(z;\tilde{\alpha}_i,\tilde{\mathrm{g}})
    }\label{eq:loop_equation_1},
\end{split}
\end{align}
where
\begin{align}
\begin{split}
\hat{f}(z;\tilde{\alpha}_i,\tilde{\mathrm{g}}_\ell)
&= g_s b_E \sum_{I=1}^N
        \frac{
                W'(z;\tilde{\alpha}_i,\mathrm{\tilde{g}}_\ell)
                -W'(\lambda_I;\tilde{\alpha}_i,\tilde{\mathrm{g}}_\ell)
        }{z-\lambda_I}\\
&=\frac{\bar{\alpha}_1+\epsilon}{z}\hat{w}(0)
+\frac{\bar{\alpha}_2}{z-q}\hat{w}(q)
+\frac{\bar{\alpha}_3}{z-1}\hat{w}(1)
+g_s b_E\sum_{I=1}^N\frac{J'(z)-J'(\lambda_I)}{z-\lambda_I}.
\end{split}
\end{align}

Free energy $F=\log Z$ and the resolvent
        are related by the operation
\begin{align}
\orddif{\ \ \ }{J(z)}
&\equiv \sum_{\ell=0}^\infty \frac{1}{(z-z_0)^{\ell+1}}\pardif{}{\tilde{\mathrm{g}}_\ell}.
\label{eq:def_loop_ins_op}
\end{align}
We obtain
\begin{align}
\orddif{\ \ }{J(z)}\Big(g_s^2 F \Big)&=\Average{\hat{w}(z)}-\frac{\varrho}{z-z_{0}}.
\end{align}
Acting once again, we obtain
\begin{align}
\begin{split}
\orddif{\ \ }{J(z_2)}\orddif{\ \ }{J(z_1)}\Big(g_s^2 F \Big)
&=\frac{1}{g_s^2}\left(
        \Average{\hat{w}(z_2)\hat{w}(z_1)}-\Average{\hat{w}(z_2)}\Average{\hat{w}(z_1)}
\right)\\
&\equiv\frac{1}{g_s^2} \Average{\prod_{i=1,2} \hat{w}(z_i)}_{\mathrm{conn}}.
\end{split}\label{eq:F''=w^2}
\end{align}
Likewise,
\begin{align}
\prod_{i=1}^n \left(
        \orddif{\ \ \ }{J(z_i)}
    \right)
    \Big( g_s^2 F\Big)
&=      \left(
                \frac{1}{g_s^2}
        \right)^{n-1}
        \Average{\prod_{i=1}^n\hat{w}(z_i)}_{\mathrm{conn}}
        -\frac{\varrho}{z-z_{0}}\delta_{n,1}.
\end{align}

Let us first make a double expansion of (\ref{eq:loop_equation_1})
        in $g_s$ and in $\epsilon$, setting $J=0$.
Let
\begin{align}
\begin{split}
\Average{\hat{w}(z)}_{J=0}\equiv w(z)
&= \sum_{i=0}^\infty g_s^i \sum_{j=1}^\infty \epsilon^j w_{i,j}(z),\\
\Average{\hat{f}(z)}_{J=0}\equiv f(z)
&= \sum_{i=0}^\infty g_s^i \sum_{j=1}^\infty \epsilon^j f_{i,j}(z),\\
W'(z;\tilde{\alpha}_i,\tilde{\mathrm{g}}_\ell)\Big|_{J=0}
&=W'_{0,0}(z)+\epsilon W'_{0,1}(z).
\end{split}
\label{eq:expand_in_g_e}
\end{align}
Below, $w_{i,j}(z)$ and $f_{i,j}(z)$ are
        also denoted by $w^{[\frac{i+j}{2}]}_j(z)$, $f^{[\frac{i+j}{2}]}_j(z)$.
Substituting
(\ref{eq:F''=w^2}) and
(\ref{eq:expand_in_g_e}) into (\ref{eq:loop_equation_1}) at $J=0$,
        we obtain
\begin{align}
\begin{split}
0&=\epsilon \sum_{i=0}^\infty g_s^i \sum_{j=0}^\infty \epsilon^j w'_{i,j}(z)
        +\left(
                \sum_{i=0}^\infty g_s^i \sum_{j=0}^\infty \epsilon^j w_{i,j}(z)
        \right)^2
        +g_s^2 \sum_{i=0}^\infty g_s^i \sum_{j=0}^\infty \epsilon^j
                \orddif{\ \ \ }{J(z)}w_{i,j}(z)\\
&\hs+\Big(
                W'_{0,0}(z)+\epsilon W'_{0,1}(z)
        \Big)
        \left(
                \sum_{i=0}^\infty g_s^i \sum_{j=0}^\infty \epsilon^j w_{i,j}(z)
        \right)
-\sum_{i=0}^\infty g_s^i \sum_{j=0}^\infty \epsilon^j f_{i,j}(z)
\end{split}
\end{align}
In the lowest order $g_s^0 \epsilon^0$,
        we have the planar result
\begin{align}
\begin{split}
\mathcal{Y}_{0,0}(z)&\equiv w_{0,0}(z)+\frac{1}{2}W'_{0,0}(z)\\
0&=\mathcal{Y}_{0,0}(z)^2-\frac{1}{4}W'_{0,0}(z)^2-f_{0,0}(z)
\end{split}
\label{eq:planar result}
\end{align}
In the order $g_s^i\epsilon^j$, we have
\begin{align}
\begin{split}
0&=w'_{i,j-1}(z)+\sum_{i_1+i_2=i}\ \sum_{j_1+j_2=j}w_{i_1,j_1}(z)w_{i_2,j_2}(z)\\
&\hs+\orddif{\ \ \ }{J(z)}w_{i-2,j}(z)\Big|_{J=0}
+W'_{0,0}(z) w_{i,j}(z)+W'_{0,1}(z) w_{i,j-1}(z)-f_{i,j}(z).
\end{split}
\label{gseij}
\end{align}
These can be treated recursively.
In particular,
\begin{align}
\begin{split}
g_s^0 \epsilon^1:&\ \ 
    0=2\mathcal{Y}_{0,0}(z)w_{0,1}(z)
    + \left(
        W'_{0,1}(z)
        +\orddif{}{z}        
    \right)
    w_{0,0}(z)
    -f_{0,1}(z),\\
g_s^1 \epsilon^0:&\ \ 
    0=2\mathcal{Y}_{0,0}(z)w_{1,0}(z)
    -f_{1,0}(z),\\
g_s^0 \epsilon^2:&\ \ 
    0=2\mathcal{Y}_{0,0}(z)w_{0,2}(z)
    +w_{0,1}(z)^2
    + \left(
        W'_{0,1}(z)
        +\orddif{}{z}        
    \right) w_{0,1}(z)
    -f_{0,2}(z),\\
g_s^1 \epsilon^1:&\ \ 
0=2\mathcal{Y}_{0,0}(z) w_{1,1}(z)
    +2w_{1,0}(z)w_{0,1}(z)
    + \left(
        W'_{0,1}(z)
        +\orddif{}{z}        
    \right) w_{1,0}(z)
    -f_{1,1}(z),\\
g_s^2 \epsilon^0:&\ \ 
    0=2\mathcal{Y}_{0,0}(z) w_{2,0}(z) + w_{1,0}(z)^2
    +\orddif{\ \ \ }{J(z)}w_{0,0}(z)\Big|_{J=0}
    -f_{2,0}(z).
\end{split}
\label{eq:loop_eq_of_lower_ij}
\end{align}
 From the general structure of  eq.(\ref{gseij}), it is easy to conclude that
 $w_{i,j}=0$ if $i$ is odd: every term in eq.(\ref{gseij}) is linear  either in 
  $w_{i^{\prime},j^{\prime}}$  for  some  $i^{\prime} < i$, $i^{\prime}$ odd and $j^{\prime} \leq j$
   or in  $w_{i,j-1}$

\begin{figure}[tb]
  \begin{center}
    \includegraphics{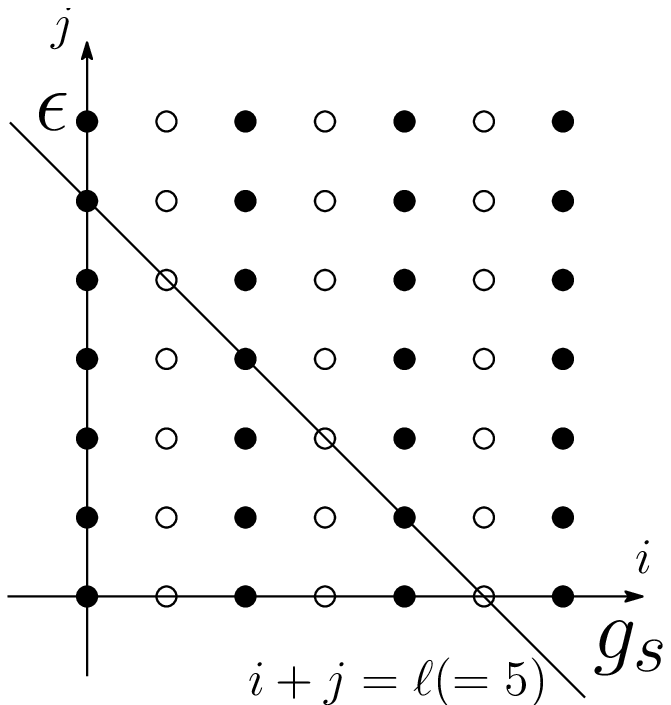}
  \end{center}
  \caption{
  }
\label{fig:lattice}
\end{figure}

Let us recall $\epsilon=Q_E g_s$ and we reorganize as the half genus expansion
\begin{align}
\begin{split}
w(z)=\sum_{\ell=0}^\infty g_s^\ell \sum_{j=0}^\ell Q_E^j \ w_{\ell-j,j}(z)
&=\sum_{\ell=0}^\infty g_s^\ell
        \sum_{j=0}^\ell Q_E^j \ w^{[\ell/2]}_j (z)\\
&\equiv \sum_{\ell=0}^\infty g_s^\ell
       \ w^{[\ell/2]} (z;Q_E).
\end{split}
\end{align}
  The sum over $j$ actually runs  over  either even or odd, depending upon
   whether $\ell$ is even  or odd.  We illustrate the original double expansion and
  the half genus expansion in Figure \ref{fig:lattice}.

As for free energy,
\begin{align}
\begin{split}
g_s^2 F=\sum_{i=0}^\infty g_s^i \sum_{j=0}^\infty \epsilon^j F_{i,j}
=\sum_{\ell=0}^\infty g_s^\ell \sum_{j=0}^\ell Q_E^j \ F_{\ell-j,j}
&=\sum_{\ell=0}^\infty g_s^\ell
        \sum_{j=0}^\ell\ Q_E^j \ F^{[\ell/2]}_j \\
&\equiv\sum_{\ell=0}^\infty g_s^\ell
       \ F^{[\ell/2]} ,
\end{split}
\end{align}
also
\begin{align}
\begin{split}
w_{i,j}(z)=\orddif{\ \ \ }{J(z)}F_{i,j}\Big|_{J=0}+\frac{\varrho}{z-z_0}\delta_{i,0}\delta_{j,0}.
\end{split}
\end{align}

\subsection{Paths}
\begin{figure}[tb]
  \begin{center}
\includegraphics{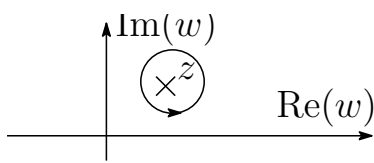}
  \end{center}
  \caption{
        {\footnotesize
                A contour $\mathcal{C}_z$.
                This contour is a small circle encircling $z$ anti-clockwise.
        }
  }
  \label{fig:contour_C_z}
\end{figure}

\begin{figure}[tb]
  \begin{center}
\includegraphics{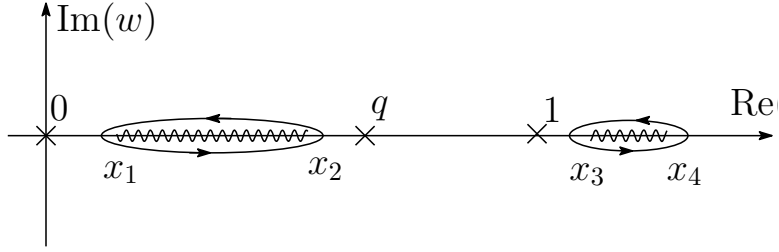}
  \end{center}
  \caption{
        {\footnotesize
                Contours $\mathcal{C}_1$ and $\mathcal{C}_2$.
                The contour $\mathcal{C}_1$ is an anti-clockwise contour
                        encircling the segment $[x_1,x_2]$.
                The contour $\mathcal{C}_2$ is an anti-clockwise contour
                        encircling the segment $[x_3,x_4]$.
                Let contour $\mathcal{C}$ be a sum of $\mathcal{C}_1$ and $\mathcal{C}_2$,
                        which encircles the cuts of $w_{0,0}(z)$.
        }
  }
\label{fig:contour_C}
\end{figure}

This subsection defines the contours of complex integrals.
Contours $C_z$, $\mathcal{C}_1$, and $\mathcal{C}_2$ are defined respectively in fig.(\ref{fig:contour_C_z}) and (\ref{fig:contour_C}), where $x_1$, $x_2$, $x_3$, and $x_4$ are the end points of the cuts of $w_{0,0}(z)$ with $x_1<x_2<x_3<x_4$.
Note that $0<x_1<x_2<q<1<x_3<x_4$.
A path $\mathcal{C}$ is sum of  $\mathcal{C}_1$ and $\mathcal{C}_2$.
The restriction $C^{(\zeta)} > C^{(\omega)}$ implies that the $\zeta$ contour lies outside
    the $\omega$ contour, both enclosing the cut  of $w_{0,0}$.
The same applies to $C^{(\omega)} > C^{(\zeta)}> C^{(\eta)}$.

\section{$F_{0,1}$ and $F_{0,2}$ }
In this section, the representation of $F_{0,1}$ and that of $F_{0,2}$ as contour integrals are presented.

\subsection{Formula for $F_{0,1}$}
\label{sec:F_01}
The formula we state is
\begin{align}
\begin{split}
F_{0,1}&=\oint_{\mathcal{C}}\frac{\de \omega}{2 \pi \im} \mathcal{Y}_{0,0}(\omega)
    \log\left[
        \omega \mathcal{Y}_{0,0}(\omega)
    \right]\\
    &=\oint_{\mathcal{C}}\frac{\de \omega}{2 \pi \im} \mathcal{Y}_{0,0}(\omega)
    \log \mathcal{Y}_{0,0}(\omega)
    +\oint_{\mathcal{C}}\frac{\de \omega}{2 \pi \im} \mathcal{Y}_{0,0}(\omega)
    \log \omega
\end{split}.
\label{eq:F_01}
\end{align}
The proof is given in appendix \ref{sec:Appendix_On_F_01}.
The point where this formula differs from that of \cite{CE0604,C1009} is the presence of the second term.

\subsection{Formula for $F_{0,2}$}
\label{sec:F_02}
The formula we state is
\begin{align}
\begin{split}
F_{0,2}
&=-\frac{1}{4}\oint_{\mathcal{C}}\frac{\de \zeta}{2\pi \im}
    \left[
        \frac{\mathcal{Y}'_{0,0}(\zeta)}{\mathcal{Y}_{0,0}(\zeta)}
        +\frac{1}{\zeta}
    \right]
    \int'_{\mathcal{C}} \frac{\de \omega}{2\pi \im}
    \frac{
        \de E_{\zeta,\bar{\zeta}}
    }{\de \omega}
    \log [\omega \mathcal{Y}_{0,0}(\omega)]\\
&\hs\hs\hs    -\frac{1}{12}
        \sum_{i=1}^4 
        \left[
                \log \abs{M_i}+\sum_{1\le j <i \le 4} \log \abs{x_i-x_j}
        \right]-\frac{1}{8}
        \sum_{i=1}^4 
                \log \abs{x_i}\\
&=-\frac{1}{4}\oint_{\mathcal{C}}\frac{\de \zeta}{2\pi \im}
        \frac{\mathcal{Y}'_{0,0}(\zeta)}{\mathcal{Y}_{0,0}(\zeta)}
    \int'_{\mathcal{C}} \frac{\de \omega}{2\pi \im}
    \frac{
        \de E_{\zeta,\bar{\zeta}}
    }{\de \omega}
    \log  \mathcal{Y}_{0,0}(\omega)-\frac{1}{12}
        \log \left(
                \abs{\Delta'(x_j)}\prod_{i=1}^4 \abs{M_i}
        \right)
        \\
&\hs\hs-\frac{1}{4}\oint_{\mathcal{C}}\frac{\de \zeta}{2\pi \im}        
        \frac{1}{\zeta}
    \int'_{\mathcal{C}} \frac{\de \omega}{2\pi \im}
    \frac{
        \de E_{\zeta,\bar{\zeta}}
    }{\de \omega}
    \log  \mathcal{Y}_{0,0}(\omega)-\frac{1}{4}\oint_{\mathcal{C}}\frac{\de \zeta}{2\pi \im}        
        \frac{\mathcal{Y}'_{0,0}(\zeta)}{\mathcal{Y}_{0,0}(\zeta)}
    \int'_{\mathcal{C}} \frac{\de \omega}{2\pi \im}
    \frac{
        \de E_{\zeta,\bar{\zeta}}
    }{\de \omega}
    \log \omega \\
&\hs\hs\hs    -\frac{1}{4}\oint_{\mathcal{C}}\frac{\de \zeta}{2\pi \im}        
        \frac{1}{\zeta}
    \int'_{\mathcal{C}} \frac{\de \omega}{2\pi \im}
    \frac{
        \de E_{\zeta,\bar{\zeta}}
    }{\de \omega}
    \log \omega-\frac{1}{8}
                \log \left(\prod_{i=1}^4\abs{x_i}\right),
\end{split}
\label{eq:F_02}
\end{align}
where the formal integral $\int'_{\mathcal{C}}\de \omega$ means
\begin{align}
\begin{split}
\int'_{\mathcal{C}}\de \omega f(\omega)
:=\left(
        \sum_{n=1}^2 \int_{x_{2n-1}}^{x_{2n}} \de \omega
\right)
\lim_{\epsilon\to +0}
        \left[
                f(\omega-\im \epsilon)-f(\omega+\im \epsilon)
        \right]
\end{split}
\end{align}
and $\Delta'(x_j)$ and $M_j$ are defined by
\begin{align}
\begin{split}
\Delta' (x_j)&\equiv\prod_{i=1}^4 \prod_{j=1} ^{i-1}(x_i-x_j),\\
M_j&\equiv\oint_{\mathcal{C}}\frac{\de\omega}{2\pi\im}
    \frac{W'_{0,0}(\omega)}{\sqrt{\prod_{i=1}^4(\omega-x_i)}}\frac{1}{\omega-x_j}.
\end{split}
\end{align}
The proof is given in appendix \ref{sec:Appendix_On_F_02}.
The point where this formula differs from that of \cite{CE0604,C1009} is the presence of the third, the fourth, the fifth, and the sixth terms.

\subsection{Representation as $\epsilon$ corrected Seiberg-Witten prepotential}
Substituting  the $0d-4d$ dictionary (\ref{eq:1d-4d_dictionary})
    into the expressions for $F_{0,1}$ and $F_{0,2}$
    in eq.(\ref{eq:F_01}) and (\ref{eq:F_02}),
    we obtain the first two $\epsilon$ corrections to the Seiberg-Witten prepotential.
In particular, we would like to convert $\prod_{i=1}^4 M_i$ and $x_i$
    into 4d parameters.
As for the former, it is accomplished by (\ref{eq:M_vs_a4_q}).
As for the latter, we show in Appendix \ref{sec:Appendix_On_Planar_limit}
    that they can in principle be expressed in terms of the 4d parameters, which requires
    solving a system of nonlinear equations.
We do not consider this in this paper.

\section*{Acknowledgements}
We thank Takeshi Oota for valuable remarks on this subject and participation
at the initial stage of this work.
The research of H.~I.~ and N.~Y.~
is supported in part by the Grant-in-Aid for Scientific Research (2054278)
as well as JSPS Bilateral Joint Projects(JSPS-RFBR collaboration)
from the Ministry of Education, Science and Culture, Japan.

\appendix
\section{Review on a few materials from Riemann surfaces}
\label{sec:Appendix_on_Riemann_surfaces}
In this section, we review some facts on the Riemann surface used in this paper. 
This review is based on
        the researches from the mathematical side\cite{R59,Fay70,FK92,Fay92,KK08}
        and
        researches of the matrix model\cite{E0407,CE0504,CE0604,C1009}.

\subsection{Hyperelliptic Riemann surface}
In this subsection, 
    we introduce definitions and notation on a hyperelliptica Riemann surface $\mathcal{M}$  
    that is given by
\begin{align}
\begin{split}
\omega^2&=\left[
        \frac{2}{\bar{\alpha}_4}z(z-q)(z-1)\mathcal{Y}_{0,0}(z)
    \right]^2\\
&=\prod_{i=1}^4 (z-x_i).
\end{split}
\end{align}
Our definitions and notation are in agreement with \cite{E0407}.

The Riemann surface $\mathcal{M}$ consists of two sheets of complex plane,
    which are joined along the cut of $\mathcal{Y}_{0,0}(z)$.
We call one of the sheets that satisfies
\begin{align}
\begin{split}
\lim_{\abs{z}\to\infty}[z\mathcal{Y}_{0,0}(z)]=+\frac{\bar{\alpha}_4}{2}
\end{split}
\end{align}
physical sheet in this paper.
The other sheet is called second sheet.
If $z$ is a local coordinate on physical sheet in Riemann surface $\mathcal{M}$,
        we denote the corresponding point on the second sheet by $\bar{z}$.
Note that
\begin{align}
\begin{split}
\mathcal{Y}_{0,0}(\bar{z})=-\mathcal{Y}_{0,0}(z),
\hs W_{0,0}(\bar{z})=W_{0,0}(z),
\hs \de \bar{z} = \de z.
\end{split}
\end{align}
The coordinates of the sheets that we regard as complex planes are
    one of the local coordinates of the local charts that are the whole sheets except the cuts.
Near the branch points $x_i$, we employ $\tau_i(z)$
        as an alternative to $z$:
\begin{align}
\begin{split}
\tau_i (z)=\sqrt{z-x_i},
\end{split}
\end{align}
which is indeed a local coordinate of the neighborhood of $x_i$.

\subsection{Bergmann kernel}
In this section, we review the Bergmann kernel based on \cite{E0407,CE0504,CE0604,C1009,Fay92,KK08}.
The Bergmann kernel, $B(z,\zeta)\de z\de \zeta$, is a symmetric meromorphic bidifferential
        on the $\mathcal{M}\times\mathcal{M}$                
        with only a double pole at $z=\zeta$
            \footnote{
                Note that $z=\bar{\zeta}$ is not a singular point.
            }
            on the whole Riemann surface
                which satisfies followings:
\begin{align}
\begin{split}
B(z,\zeta)\de z\de \zeta &\approx \frac{\de z\de\zeta}{(z-\zeta)^2}+\textrm{finite}\hs\hs(z\approx\zeta),\\
0&=\oint_{\mathcal{C}_i}B(z,\zeta)\de z\hs (i=1,2).
\end{split}
\label{eq:def_Bergmann}
\end{align}
Since $B(z,\zeta)\de z \de \zeta$ is a meromorphic bidifferential, we have
\begin{align}
\begin{split}
B(z,\zeta)
&=B(z,\tau_i(\zeta))        
        \orddif{\tau_i(\zeta)}{\zeta}\\
&=B(\tau_i(z),\zeta)
        \orddif{\tau_i(z)}{z} 
=B(\tau_i(z),\tau_i(\zeta))
        \orddif{\tau_i(z)}{z}
        \orddif{\tau_i(\zeta)}{\zeta}.
\end{split}
\label{eq:Bergmann_is_bidifferential}
\end{align}
By using (\ref{eq:Bergmann_is_bidifferential}), $B(z,\zeta)$ can be written as
\begin{align}
\begin{split}
B(z,\zeta)&=B(\tau_i(z),\zeta)\orddif{\tau_i(z)}{z}\\
&=B(\tau_i(z),\zeta)\frac{1}{2\sqrt{z-x_i}}.
\end{split}
\end{align}
Since $B(\tau_i(z),\zeta)$ is not singular at $z=x_i$,
        we can define $B([x_{i}],\zeta)$ as
\begin{align} 
\begin{split}
B([x_{i}],\zeta):&=\lim_{z\to x_i} \sqrt{z-x_i}B(z,\zeta)\\
&=\frac{1}{2}B(\tau_i(x_{i}),\zeta).
\end{split}
\end{align}
Note that $B(z,\zeta)$ is allowed to have a singularity at $z=x_i$
        since $z$ is not a local coordinate near the cut.

We introduce $\displaystyle{\orddif{E_{\omega,\omega_0}}{z}}$ as a definite integral of $B(z,\zeta)$:
\begin{align}
\begin{split}
\orddif{E_{\omega,\omega_0}}{z}=\int^\omega_{\omega_0}B(z,\zeta)\de \zeta.
\end{split}
\label{eq:dE_vs_B}
\end{align}
We obtain
\begin{align}
\begin{split}
\pardif{}{\omega}\orddif{E_{\omega,\bar{\omega}}}{z}
&=[B(z,\omega)-B(z,\bar{\omega})].
\end{split}
\label{eq:dE/dz_vs_dy/dj}
\end{align}        

The two point resolvent correlator $w_{0,0}(z,\zeta)$ and the Bergmann kernel are
    related by the following identification
\begin{align}
\begin{split}
\orddif{\mathcal{Y}_{0,0}(\zeta)}{J(z)}&= w_{0,0}(z,\zeta)+\frac{1}{2(z-\zeta)^2}\\
&=\frac{1}{2}\left[
        B(z,\zeta)-B(z,\bar{\zeta})
\right],
\end{split}
\label{eq:dy/dj_vs_B}
\end{align}
where first equality follows from (\ref{eq:planar result}).
The Bergmann kernel satisfies
\begin{align}
\begin{split}
B(z,\zeta)+B(z,\bar{\zeta})&=\frac{1}{(z-\zeta)^2},\\
B(\bar{z},\zeta)-B(z,\bar{\zeta})&=0.
\end{split}
\label{eq:Bergmann_identity}
\end{align}
By using (\ref{eq:Bergmann_identity}), we have
\begin{align}
\begin{split}
\orddif{\mathcal{Y}_{0,0}(\zeta)}{J(z)}
&=B(z,\zeta)-\frac{1}{(z-\zeta)^2}\\
&=-B(z,\bar{\zeta}).
\end{split}
\end{align}


\subsection{$\displaystyle{\pardif{}{x_i}B(z,\zeta)}$}    
Now, we show
\begin{align}
\begin{split}
\pardif{}{x_i}B(z,\zeta)=2B(z,[x_i])B([x_i],\zeta)
\end{split}
\label{eq:Rauch_Formula}
\end{align}
by using the demonstration technique of Rauch variational formula\cite{R59,Fay92}.
Our proof is based on \cite{KK08}.

First, we fix $\zeta$.
Since $B(\tau_i(z),\zeta)$ is not singular at $z=x_i$,
        we can Taylor-expand it around $z=x_i$.
We, thus, define expansion coefficients $a_n^{(i)}(\zeta)$ by
\begin{align}
\begin{split}
B(z,\zeta)
&=B(\tau_i(z),\zeta)\orddif{\tau_i(z)}{z}\\
&\equiv\frac{1}{2\tau_i(z)}
\left[
        \sum_{n=0}^\infty a_n^{(i)}(\zeta) \tau_i(z)^n
\right].
\end{split}
\label{eq:def_a(zeta)}
\end{align}        
We also define expansion coefficients $a(z_0;\zeta)$ by 
\begin{align}
\begin{split}
B(z,\zeta)
&\equiv
\left\{
\begin{array}{lr}
        \displaystyle{\sum_{n=0}^\infty a_n(z_0;\zeta)(z-z_0)^n}&\hs(z_0\neq\zeta)\\
        \displaystyle{\frac{1}{(z-\zeta)^2}+\sum_{n=0}^\infty a_n(\zeta;\zeta)(z-\zeta)^n}&\hs(z_0=\zeta)
\end{array}
\right.
.
\end{split}
\end{align}
The $x_i$-derivative of $B(z,\zeta)$ is obtained by
\begin{align*}
\pardif{B(z,\zeta)}{x_i}=
\left\{
\begin{array}{lr}
\displaystyle{\orddif{\tau_i(z)}{z}
        \Bigg[
                \frac{1}{2\tau_i(z)^2}\sum_{n=0}^\infty a_n^{(i)}(\zeta)(1-n)\tau_i(z)^n
                +\sum_{n=0}^\infty \pardif{a_n^{(i)}(\zeta)}{x_i}\tau_i(z)^n
        \Bigg]}
&\displaystyle{(z\approx x_i)}\\
\displaystyle{\orddif{\tau_j(z)}{z}
        \Bigg[
                \sum_{n=0}^\infty \pardif{a_n^{(j)}(\zeta)}{x_i}\tau_j(z)^n
        \Bigg]}
&\displaystyle{(z\approx x_j)}\\
\displaystyle{\sum_{n=0}^\infty \pardif{a_n(z_0;\zeta)}{x_i}(z-z_0)^n}
&\displaystyle{(\textrm{otherwise})}\\
\end{array}
\right.
.
\end{align*}
This implies that
        $x_i$-derivative of $B(z,\zeta)$ has
        only a double pole at $z=x_i$ with no residue.
Now, we find that
\begin{align*}
\begin{split}
\orddif{z}{\tau_i(z)}\pardif{B(z,\zeta)}{x_i}-\frac{a_0^{(i)}(\zeta)}{2\tau_i(z)^2},
\end{split}
\end{align*}
has no pole on $\mathcal{M}$.
In addition, $B(x_i,\zeta)=0$.
Since a holomorphic
function with no pole on the whole Riemann surface must be a constant,
        we have
\begin{align}
\begin{split}
\pardif{B(z,\zeta)}{x_i}=\frac{a_0^{(i)}(\zeta)}{2\tau_i(z)^2}\orddif{\tau_i(z)}{z}.
\end{split}
\label{eq:db/dx=a(zeta)f(tau(z))}
\end{align}
Repeating the same argument for $\zeta$, we obtain
\begin{align}
\begin{split}
\pardif{B(z,\zeta)}{x_i}=\orddif{\tau_i(\zeta)}{\zeta}\frac{a_0^{(i)}(z)}{2\tau_i(\zeta)^2}.
\end{split}
\label{eq:db/dx=a(z)f(tau(zeta))}
\end{align}
Comparing (\ref{eq:db/dx=a(zeta)f(tau(z))}) and (\ref{eq:db/dx=a(z)f(tau(zeta))}),
        we have
\begin{align}
\begin{split}
a_0^{(i)}(z)=\frac{C}{\tau_i(z)^2}\orddif{\tau_i(z)}{z},
\end{split}
\end{align}
where $C$ is a constant.

In order to determine $C$, we investigate behavior of $B(\tau_i(z),\tau(x_i))$ in the neighborhood of $z=x_i$.
Estimating the lowest order of $\tau(x_i)$
        by using (\ref{eq:Bergmann_is_bidifferential}) and (\ref{eq:def_a(zeta)}),
        we have
\begin{align}
\begin{split}
B(\tau_i(z),\tau(x_i))
&=B(z,\zeta)\orddif{z}{\tau_i(z)}
\orddif{\zeta}{\tau_i(\zeta)}\bigg|_{\zeta=x_i}\\
&=
a_0^{(i)}(z)
\orddif{\tau_i(\zeta)}{\zeta}
\orddif{z}{\tau_i(z)}
\orddif{\zeta}{\tau_i(\zeta)}\bigg|_{\zeta=x_i}
=\frac{C}{\tau_i(z)^2}.
\end{split}
\end{align}        
Eq.(\ref{eq:def_Bergmann}) and (\ref{eq:Bergmann_is_bidifferential}) imply
\begin{align}
\begin{split}
B(\tau_i(z),\tau(x_i))
&\approx
\frac{1}{(\tau_i(z)-\tau(x_i))^2} + \textrm{finite}
=\frac{1}{\tau_i(z)^2}+ \textrm{finite}.
\end{split}
\end{align}
We thus obtain $C=1$.

By using eq.(\ref{eq:def_a(zeta)}), we denote $a_0^{(i)}(\zeta)$ as
\begin{align}
\begin{split}
a_0^{(i)}(\zeta)=\lim_{z\to x_i} 2 \tau_i(z)B(z,\zeta)=2B([x_i],\zeta).
\end{split}
\end{align}
We thus obtain
\begin{align*}
\begin{split}
\pardif{B(z,\zeta)}{x_i}
=\frac{a_0^{(i)}(\zeta)}{2\tau_i(z)^2}\orddif{\tau_i(z)}{z}
=\frac{a_0^{(i)}(z)a_0^{(i)}(\zeta)}{2}
=2B(z,[x_i])B([x_i],\zeta).
\end{split}
\end{align*}
Eq.(\ref{eq:Rauch_Formula}) has been shown.

\subsection{
                $\displaystyle{\orddif{\ \ }{J(z)}B(\eta,\zeta)}$
                and
                $\displaystyle{\orddif{\ \ }{J(z)}\orddif{E_{\zeta,\bar{\zeta}}}{\omega}}$
}
To obtain $\displaystyle{\orddif{B(\eta,\zeta)}{J(z)}}$
                and
                $\displaystyle{\orddif{\ \ }{J(z)}\orddif{E_{\zeta,\bar{\zeta}}}{\omega}}$,
        we start to study $\displaystyle{\orddif{x_i}{J(z)}}$.
        The loop insertion operator acts on $\mathcal{Y}_{0,0}(z)$ as
\begin{align}
\begin{split}
\orddif{\mathcal{Y}_{0,0}(\zeta)}{J(z)}
&=\orddif{x_{i}}{J(z)}\pardif{\mathcal{Y}_{0,0}(\zeta)}{x_i}
        =-\frac{1}{2}\sum_{i=1}^4\frac{1}{\zeta-x_i}\orddif{x_{i}}{J(z)}\mathcal{Y}_{0,0}(\zeta)\\ &=\frac{1}{2}\left[B(z,\zeta)-B(\bar{z},\zeta)\right],
\end{split}
\label{eq:dy/dj_vs_x_i}
\end{align}
where eq.(\ref{eq:dy/dj_vs_B}) have been used in the second line.
Eq.(\ref{eq:dy/dj_vs_x_i}) implies 
\begin{align}
\begin{split}
\orddif{x_i}{J(z)}&=-\lim_{\zeta\to x_i}
        \frac{
                \left[
                        B(z,\zeta)-B(\bar{z},\zeta)
                \right]
                (\zeta-x_i)
        }{\mathcal{Y}_{0,0}(\zeta)}\\
&=-
        \frac{
                \left[
                        B(z,[x_i])-B(\bar{z},[x_i])
                \right]  
        }{\mathcal{Y}_{0,0}([x_i])},
\end{split}
\label{eq:dx/dj_vs_2b/y}
\end{align}
where $\mathcal{Y}_{0,0}([x_i])$ is defined by
\begin{align}
\begin{split}
\mathcal{Y}_{0,0}(z)=
\frac{\mathcal{Y}_{0,0}([x_{i}])}{\sqrt{z-x_i}}+\mathcal{O}[\sqrt{z-x_i}],
\end{split}
\label{eq:y_approx_f(eta-x)}
\end{align}
in other words, $\mathcal{Y}_{0,0}([x_i])$ is determined by
\begin{align*}
\begin{split}
\mathcal{Y}_{0,0}([x_i])=
\lim_{z\to x_i}\mathcal{Y}_{0,0}(z)\sqrt{z-x_i}.
\end{split}
\end{align*}

Since $B(z,\zeta)$ is determined by $x_i$ completely,
        $B(z,\zeta)$ only depend on $J(z)$ through $x_i$.
we thus obtain
\begin{align}
\begin{split}
\orddif{B(\omega,\zeta)}{J(z)}=w_{0,0}(z,\omega,\zeta)
&=\sum_{i=1}^4 \pardif{B(\omega,\zeta)}{x_i} \orddif{x_i}{J(z)}\\
&=2\sum_{i=1}^4 B(\omega,[x_i])B([x_i],\zeta) \orddif{x_i}{J(z)}\\
&=-2\sum_{i=1}^4  
        \frac{
                B(\omega,[x_i])B([x_i],\zeta)\left[
                        B(z,[x_i])-B(\bar{z},[x_i])
                \right]  
        }{\mathcal{Y}_{0,0}([x_i])}\\
&=-2\sum_{i=1}^4  
        \frac{
                B(\zeta,[x_i])\left[
                        B(\omega,[x_i])-B(\bar{\omega},[x_i])
                \right]B(z,[x_i])  
        }{\mathcal{Y}_{0,0}([x_i])},
\end{split}
\label{eq:db/dj}
\end{align}
where we have used eq.(\ref{eq:Rauch_Formula})
in the second equality and eq.(\ref{eq:dx/dj_vs_2b/y}) in the third.
The fourth equality holds
        because $w_{0,0}(z,\omega,\zeta)$ is symmetric for $z$, $\omega$, and $\zeta$.

If $\eta \approx x_i$, we have
\begin{align}
\begin{split}
\orddif{E_{\eta,\bar{\eta}}}{z}
&=\int^\eta_{\bar{\eta}}
    \left[
        \frac{B(z,[x_i])}{\sqrt{\zeta-x_i}}
        +\mathcal{O}[\sqrt{\zeta-x_i}]
    \right]\de\zeta
=\int^{\tau_i(\eta)}_{\tau_i(\bar{\eta})}
    \left[
        2B(z,[x_i])
        \frac{}{{}}
        +\mathcal{O}[\tau_i(\zeta)]
    \right]\de\tau_i(\zeta)\\
&=4B(z,[x_i])\sqrt{\eta-x_i}+\mathcal{O}[(\eta-x_i)^{3/2}].
\end{split}
\label{eq:de/dz_approx_f(eta-x)}
\end{align}
We can denote (\ref{eq:db/dj}) as
\begin{align}
\begin{split}
\orddif{B(\omega,\zeta)}{J(z)}
&=-\sum_{i=1}^4 \Res_{\eta=x_i}
        \frac{
                B(\zeta,\eta)\left[
                        B(\omega,\eta)-B(\omega,\bar{\eta})
                \right]  
        }{2\mathcal{Y}_{0,0}(\eta)}
        \orddif{E_{\eta,\bar{\eta}}}{z}.
\end{split}
\label{eq:db/dj_vs_res}
\end{align}
Integrating eq.(\ref{eq:db/dj_vs_res}) with respect to $\zeta$,
        we have
\begin{align}
\begin{split}
\orddif{\ \ }{J(z)}\orddif{E_{\zeta,\bar{\zeta}}}{\omega}=-\sum_{i=1}^4 \Res_{\eta=x_i}
                \frac{1}{2\mathcal{Y}_{0,0}(\eta)}
        \orddif{E_{\zeta,\bar{\zeta}}}{\eta}
        \orddif{E_{\eta,\bar{\eta}}}{z}
        \left[
            B(\omega,\eta)-B(\omega,\bar{\eta})
        \right].
\end{split}
\label{eq:dE/dj_vs_res}
\end{align}

\subsection{Integral formulas}
In this subsection, we introduce two integral formulas
        that have been shown in \cite{CE0504,CE0604}. 
The first formula is
\begin{align}
\begin{split}
\oint_{\mathcal{C}}\frac{\de \omega}{2\pi\im}
        \orddif{
                E_{\omega,\bar{\omega}}
        }{z}
        w_{i,j}(\omega)=w_{i,j}(z).
\end{split}
\label{eq:int_E_w=w}
\end{align}
Since $w_{i,j}(\omega)$ is regular outside the contour
        and $\mathcal{O}[1/z^2]$ at the limit $z \to \infty$
    and $B(z,\zeta)$ behaves as
\begin{align*}
\begin{split}
\orddif{E_{\omega,\bar{\omega}}}{z}
&=\int^\omega_{\omega_0}
        B(z,\zeta)\de\zeta+\int^{\omega_0}_{\bar{\omega}}
        B(z,\zeta)\de\zeta\\
&=\int^\omega_{\omega_0}
        \left[
                \frac{1}{(z-\zeta)^2}
                +\mathcal{O}[1]
        \right]\de\zeta+\int^{\omega_0}_{\bar{\omega}}
        \mathcal{O}[1]\de\zeta\\
&\approx \frac{1}{z-\omega} \hs (\omega \approx z),
\end{split}
\end{align*}
we have
\begin{align*}
\begin{split}
\oint_{\mathcal{C}}\frac{\de \omega}{2\pi\im}
        \orddif{
                E_{\omega,\bar{\omega}}
        }{z}
        w_{i,j}(\omega)=-\oint_{\mathcal{C}_z}\frac{\de \omega}{2\pi\im}
         \frac{1}{z-\omega}
        w_{i,j}(\omega) =w_{i,j}(z).
\end{split}
\end{align*}

Second formula is
\begin{align}
\begin{split}
\oint_{\mathcal{C}}\frac{\de \omega}{2\pi\im}
        \frac{1}{\mathcal{Y}_{0,0}(\omega)}
        \orddif{
                E_{\omega,\bar{\omega}}
        }{z}
        f(\omega)=0,
\end{split}
\label{eq:int_E/2y_f=0}
\end{align}
where $f(\omega)$ is regular in the region encircled by $\mathcal{C}$.
By using eq.(\ref{eq:y_approx_f(eta-x)})
        and
        (\ref{eq:de/dz_approx_f(eta-x)}), we obtain
\begin{align*}
\begin{split}
\frac{1}{\mathcal{Y}_{0,0}(\omega)}
        \orddif{
                E_{\omega,\bar{\omega}}
        }{z}
&=\frac{4B(z,[x_i])}{\mathcal{Y}_{0,0}([x_i])}+\mathcal{O}[\omega-x_i].
\end{split}
\end{align*}
This implies integrand of eq.(\ref{eq:int_E/2y_f=0}) is regular
        in the region encircled by $\mathcal{C}$. Eq.(\ref{eq:int_E/2y_f=0})
has been shown.

\section{
        Proof of  
        $\orddif{\ \ \ }{J(z)}F_{0,1}\Big|_{J=0} = w_{0,1}(z)$
}
\label{sec:Appendix_On_F_01}
In this appendix, we confirm eq.(\ref{eq:F_01}).
In order to show this, we show the following formula in the first place: 
\begin{align}
\begin{split}
 \frac{}{} \oint_{\mathcal{C}_i} \frac{w(\omega)\de\omega}{2\pi\im}
&=\varrho\left[
        \frac{N_L}{N}\delta_{i,1}+\frac{N_R}{N}\delta_{i,2}
\right].
\end{split}
\label{eq:DV_prescription}
\end{align}
In the case of $i=1$, this is shown as
\begin{align}
\begin{split}
\oint_{\mathcal{C}_1} \frac{w(\omega)\de\omega}{2\pi\im}
&=\frac{\varrho}{N}\oint_{\mathcal{C}[x_1,x_2]} \Average{\sum_{I=1}^N\frac{1}{\omega-\lambda_I}}\frac{\de\omega}{2\pi\im}\\
&=\frac{\varrho}{N}\lim_{\varepsilon\to 0}\int^{q}_{0} \Average{\sum_{I=1}^{N}
                        \left[
                                \frac{1}{\omega-\lambda_I-\im \varepsilon}
                                -\frac{1}{\omega-\lambda_I+\im \varepsilon}
                        \right]
                    }\frac{\de\omega}{2\pi\im}
\\
&=\frac{\varrho}{N}\int^{q}_{0} \Average{\sum_{I=1}^{N}
                        \delta(\omega-\lambda_I)
                    }\de\omega\\
&=\frac{\varrho}{N}\int^{q}_{0} \Average{\sum_{I=1}^{N_{L}}
                        \delta(\omega-\lambda_I)
                    }\de\omega\\
&=\frac{ N_L}{N}\varrho,
\end{split}
\end{align}
where we have used
\begin{align*}
\begin{array}{rll}
0<x_1<&x_2 < q < 1 < x_3 <x_4,&\\
&0<\lambda_I<q\ \ &\hs(1\le I \le N_L),\\
&1<\lambda_I\ \ &\hs(N_L+1 \le I \le N).
\end{array}
\end{align*}
In the case of $i=2$, eq.(\ref{eq:DV_prescription}) is shown by the same argument.
Expanding the double expansion of $w(z)$,
        in $g_s$ and in $Q_E$, we have
\begin{align}
\begin{array}{rlr}
\displaystyle{\oint_{\mathcal{C}_k} \frac{w_{0,0}(\omega)\de\omega}{2\pi\im}}
&\displaystyle{=\varrho\left[
        \frac{N_L}{N}\delta_{k,1}+\frac{N_R}{N}\delta_{k,2}
        \right]},&\\
\displaystyle{\oint_{\mathcal{C}_k} \frac{w_{i,j}(\omega)\de\omega}{2\pi\im}}
&\displaystyle{=0}
        &\displaystyle{\hs(i>0\  \mathrm{or}\ j>0)}.
\end{array}
\label{eq:DV_prescription_with_expansion}
\end{align}

Now, we calculate $\orddif{\ \ \ }{J(z)}F_{0,1}$ as
\begin{align}
\begin{split}
\orddif{\ \ \ }{J(z)}F_{0,1}&=
\orddif{\ \ \ }{J(z)}
    \oint_{\mathcal{C}}\frac{\de \omega}{2 \pi \im}
    \mathcal{Y}_{0,0}(\omega)
    \log\left[
        \omega \mathcal{Y}_{0,0}(\omega)
    \right]\\
&=
    \oint_{\mathcal{C}}\frac{\de \omega}{2 \pi \im}
    \left\{
        \orddif{\mathcal{Y}_{0,0}(\omega)}{J(z)}
        \log\left[
                \omega \mathcal{Y}_{0,0}(\omega)
        \right]
        +
        \orddif{\mathcal{Y}_{0,0}(\omega)}{J(z)}
    \right\}\\
&=
    \oint_{\mathcal{C}}\frac{\de \omega}{2 \pi \im}
        \orddif{\mathcal{Y}_{0,0}(\omega)}{J(z)}
        \log\left[
                \omega \mathcal{Y}_{0,0}(\omega)
        \right]\\
&=
    -\oint_{\mathcal{C}}\frac{\de \omega}{2 \pi \im}
        \orddif{E_{\omega,\bar{\omega}}}{z}
        \frac{1}{2}
        \left[
                \frac{\mathcal{Y}'_{0,0}(\omega)}{\mathcal{Y}_{0,0}(\omega)}
                +\frac{1}{\omega}
        \right]\\
&=
    -\oint_{\mathcal{C}}\frac{\de \omega}{2 \pi \im}
        \orddif{E_{\omega,\bar{\omega}}}{z}
        \frac{1}{2\mathcal{Y}_{0,0}(\omega)}
        \left[
                w'_{0,0}(\omega)
                +\frac{1}{\omega}w_{0,0}(\omega)-\frac{1}{\omega}w_{0,0}(0)
        \right]\\
&=
    \oint_{\mathcal{C}}\frac{\de \omega}{2 \pi \im}
        \orddif{E_{\omega\bar{,\omega}}}{z}
        \frac{1}{2\mathcal{Y}_{0,0}(\omega)}
        \left[
                2\mathcal{Y}_{0,0}(\omega)w_{0,1}(\omega)        
        \right]\\
&=
                w_{0,1}(\omega).        
\end{split}
\end{align}
In the third equality, we have used
\begin{align}
\begin{split}
\orddif{\ \ }{J(z)}\oint_{\mathcal{C}}\frac{\de \omega}{2 \pi \im}\mathcal{Y}_{0,0}(\omega)
=\orddif{\ \ }{J(z)}\oint_{\mathcal{C}}\frac{\de \omega}{2 \pi \im}w_{0,0}(\omega)=\orddif{\ \ }{J(z)}\varrho=0,
\end{split}
\end{align}
which is shown by the eq.(\ref{eq:DV_prescription_with_expansion}).
The fifth equality holds because of the eq.(\ref{eq:int_E/2y_f=0}).
The sixth equality holds due to the loop equation (\ref{eq:loop_eq_of_lower_ij}).

\section{
        Proof of 
        $\orddif{\ \ \ }{J(z)}F_{0,2}\Big|_{J=0} = w_{0,2}(z)$
}
\label{sec:Appendix_On_F_02}
In this appendix, we confirm eq.$(\ref{eq:F_02})$.
We define $F_{0,2}^{(0)}$, $F_{0,2}^{(1)}$, and $F_{0,2}^{(2)}$ as
\begin{align}
\begin{split}
F_{0,2}^{(0)}:&=-\frac{1}{4}\oint_{\mathcal{C}}\frac{\de \zeta}{2\pi\im}
        \left[
                \frac{\mathcal{Y}'_{0,0}(\zeta)}{\mathcal{Y}_{0,0}(\zeta)}
                +\frac{1}{\zeta}
        \right]
        \int'_{\mathcal{C}}\frac{\de \omega}{2\pi\im}
                \orddif{E_{\zeta,\bar{\zeta}}}{\omega}
                \log \left[
                        \omega \mathcal{Y}_{0,0}(\omega)\right],\\
F_{0,2}^{(1)}:&=-\frac{1}{8}
        \sum_{i=1}^4 
                \log \abs{x_i}\\
F_{0,2}^{(2)}:&=-\frac{1}{12}
        \sum_{i=1}^4 
        \left[
                \log \abs{M_i}+\sum_{1\le j <i \le 4} \log \abs{x_i-x_j}
        \right],
\end{split}
\end{align}
respectively.
This definition implies
\begin{align}
\begin{split}
F_{0,2}=F_{0,2}^{(0)}+F_{0,2}^{(\textrm{1})}+F_{0,2}^{(\textrm{2})}.
\end{split}
\end{align}
In the following subsection,
    we compute $\displaystyle{\orddif{F_{0,2}^{(i)}}{J(z)}\ \ (i=1,2,3)}$
    and confirm that the sum of them is $w_{0,2}$.
    
\subsection{On $\displaystyle{\orddif{F_{0,2}^{(0)}}{J(z)}}$}
$\displaystyle{\orddif{F_{0,2}^{(0)}}{J(z)}}$ is given by
\begin{align}
\begin{split}
\orddif{F_{0,2}^{(0)}}{J(z)}&=w_{0,2}^{(a)}(z)+w_{0,2}^{(b)}(z)+w_{0,2}^{(c)}(z),
\end{split}
\end{align}
where
\begin{align}
\begin{split}
w_{0,2}^{(a)}(z)
:&=-\frac{1}{4}\oint_{\mathcal{C}}\frac{\de \zeta}{2\pi\im}
        \orddif{\ \ \ }{J(z)}\left[
                \frac{\mathcal{Y}'_{0,0}(\zeta)}{\mathcal{Y}_{0,0}(\zeta)}
                +\frac{1}{\zeta}
        \right]
        \int'_{\mathcal{C}}\frac{\de \omega}{2\pi\im}
                \orddif{E_{\zeta,\bar{\zeta}}}{\omega}
                \log \left[
                        \omega \mathcal{Y}_{0,0}(\omega)
                \right],\\
w_{0,2}^{(b)}(z)
:&=-\frac{1}{4}\oint_{\mathcal{C}}\frac{\de \zeta}{2\pi\im}
        \left[
                \frac{\mathcal{Y}'_{0,0}(\zeta)}{\mathcal{Y}_{0,0}(\zeta)}
                +\frac{1}{\zeta}
        \right]
        \int'_{\mathcal{C}}\frac{\de \omega}{2\pi\im}
                \orddif{E_{\zeta,\bar{\zeta}}}{\omega}
                \orddif{\ \ \ }{J(z)}\log \left[
                        \omega \mathcal{Y}_{0,0}(\omega)
                \right],\\
w_{0,2}^{(c)}(z)
:&=-\frac{1}{4}\oint_{\mathcal{C}}\frac{\de \zeta}{2\pi\im}
        \left[
                \frac{\mathcal{Y}'_{0,0}(\zeta)}{\mathcal{Y}_{0,0}(\zeta)}
                +\frac{1}{\zeta}
        \right]
        \int'_{\mathcal{C}}\frac{\de \omega}{2\pi\im}
                \orddif{\ \ \ }{J(z)}
                \left[
                        \orddif{E_{\zeta,\bar{\zeta}}}{\omega}
                \right]
                \log \left[
                        \omega \mathcal{Y}_{0,0}(\omega)
                \right].
\end{split}
\end{align}
In the following three subsubsections,
    we compute $w_{0,2}^{(a)}(z)$, $w_{0,2}^{(b)}(z)$, and $w_{0,2}^{(c)}(z)$, respectively.
\subsubsection{$w_{0,2}^{(a)}$}
$w_{0,2}^{(a)}(z)$ is given by
\begin{align}
\begin{split}
w_{0,2}^{(a)}(z)&=-\frac{1}{4}\oint_{\mathcal{C}}\frac{\de \zeta}{2\pi\im}
        \pardif{}{\zeta}\left[
                \orddif{\ \ \ }{J(z)}
                \log \mathcal{Y}_{0,0}(\zeta)
        \right]
        \int'_{\mathcal{C}}\frac{\de\omega}{2\pi\im}
                \orddif{E_{\zeta,\bar{\zeta}}}{\omega}
                \log \left[
                        \omega \mathcal{Y}_{0,0}(\omega)
                \right]\\
&=\frac{1}{4}\oint_{\mathcal{C}}\frac{\de \zeta}{2\pi\im}
        \frac{1}{2\mathcal{Y}_{0,0}(\zeta)}
        \left[
                B(z,\zeta)-B(z,\bar{\zeta})
        \right]
        \int'_{\mathcal{C}}\frac{\de \omega}{2\pi\im}
                \left[
                B(\omega,\zeta)-B(\omega,\bar{\zeta})
        \right]
        \log[\omega\mathcal{Y}_{0,0}(\omega)]\\
&=-\frac{1}{4}\oint_{\mathcal{C}}\frac{\de \zeta}{2\pi\im}
        \frac{1}{2\mathcal{Y}_{0,0}(\zeta)}
            \orddif{E_{\omega,\bar{\omega}}}{\zeta}
        \int'_{\mathcal{C}}\frac{\de \omega}{2\pi\im}
                \left[
                B(\omega,\zeta)-B(\omega,\bar{\zeta})
        \right]
            \left[
                    \frac{\mathcal{Y}'_{0,0}(\omega)}{\mathcal{Y}_{0,0}(\omega)}
                    +\frac{1}{\omega}
            \right]\\
&=\frac{1}{4}\mathop{\oint\oint}_{\mathcal{C}^{(\zeta)}>\mathcal{C}^{(\omega)}}
        \frac{\de\zeta}{2\pi\im}\frac{\de\omega}{2\pi\im}
                \orddif{E_{\zeta,\bar{\zeta}}}{z}
        \pardif{}{\zeta}
        \left[
            \frac{1}{2\mathcal{Y}_{0,0}(\zeta)}
            \orddif{E_{\omega,\bar{\omega}}}{\zeta}
        \right]
            \left[
                    \frac{\mathcal{Y}'_{0,0}(\omega)}{\mathcal{Y}_{0,0}(\omega)}
                    +\frac{1}{\omega}
            \right]\\
&=\frac{1}{4}\mathop{\oint\oint}_{\mathcal{C}^{(\zeta)}>\mathcal{C}^{(\omega)}}
        \frac{\de\zeta}{2\pi\im}\frac{\de\omega}{2\pi\im}
        \frac{1}{2\mathcal{Y}_{0,0}(\zeta)}
                \orddif{E_{\zeta,\bar{\zeta}}}{z}
        \left(
                \pardif{}{\zeta}-\frac{\mathcal{Y}'_{0,0}(\zeta)}{\mathcal{Y}_{0,0}(\zeta)}
        \right)
        \orddif{E_{\omega,\bar{\omega}}}{\zeta}
            \left[
                    \frac{\mathcal{Y}'_{0,0}(\omega)}{\mathcal{Y}_{0,0}(\omega)}
                    +\frac{1}{\omega}
            \right]\\
&=-\frac{1}{2}\oint_{\mathcal{C}}
        \frac{\de\zeta}{2\pi\im}
        \frac{1}{2\mathcal{Y}_{0,0}(\zeta)}
                \orddif{E_{\zeta,\bar{\zeta}}}{ z}
        \left[
                \pardif{}{\zeta}-\frac{\mathcal{Y}'_{0,0}(\zeta)}{\mathcal{Y}_{0,0}(\zeta)}
        \right]
                w_{0,1}(\zeta).
\end{split}
\end{align}
\subsubsection{$w_{0,2}^{(b)}$}
$w_{0,2}^{(b)}(z)$ is given by
\begin{align}
\begin{split}
w_{0,2}^{(b)}(z)&=-\frac{1}{4}\oint_{\mathcal{C}}\frac{\de \zeta}{2\pi\im}
        \left[
                \frac{\mathcal{Y}'_{0,0}(\zeta)}{\mathcal{Y}_{0,0}(\zeta)}
                +\frac{1}{\zeta}
        \right]
        \int'_{\mathcal{C}}\frac{\de \omega}{2\pi\im}
                \orddif{E_{\zeta,\bar{\zeta}}}{\omega}
                \frac{1}{\mathcal{Y}_{0,0}(\omega)}
                \orddif{\mathcal{Y}_{0,0}(\omega)}{J(z)}\\
&=-\frac{1}{4}\mathop{\oint\oint}_{\mathcal{C}^{(\zeta)}>\mathcal{C}^{(\omega)}}
        \frac{\de \zeta}{2\pi\im}\frac{\de \omega}{2\pi\im}
        \left[
                \frac{\mathcal{Y}'_{0,0}(\zeta)}{\mathcal{Y}_{0,0}(\zeta)}
                +\frac{1}{\zeta}
        \right]
                \orddif{\mathcal{Y}_{0,0}(\omega)}{J(z)}
                                \orddif{E_{\zeta,\bar{\zeta}}}{\omega}
                \frac{1}{\mathcal{Y}_{0,0}(\omega)}.
\end{split}
\end{align}
By using an identity 
\begin{align}
\begin{split}
\mathop{\oint\oint}_{\mathcal{C}^{(\zeta)}>\mathcal{C}^{(\omega)}}
        \frac{\de \zeta}{2\pi\im}\frac{\de \omega}{2\pi\im}\mathfrak{f}(\omega, \zeta)
\equiv \mathop{\oint\oint}_{\mathcal{C}^{(\omega)}>\mathcal{C}^{(\zeta)}}
        \frac{\de \zeta}{2\pi\im}\frac{\de \omega}{2\pi\im}\mathfrak{f}(\omega, \zeta)
+\oint_{\mathcal{C}_\omega}\frac{\de \zeta}{2\pi\im}
        \oint_{\mathcal{C}}
        \frac{\de \omega}{2\pi\im}\mathfrak{f}(\omega, \zeta),
\end{split}
\end{align}
we have
\begin{align}
\begin{split}
w_{0,2}^{(b)}(z)&=-\frac{1}{4}\mathop{\oint\oint}_{\mathcal{C}^{(\omega)}>\mathcal{C}^{(\zeta)}}
        \frac{\de \zeta}{2\pi\im}\frac{\de \omega}{2\pi\im}
        \left[
                \frac{\mathcal{Y}'_{0,0}(\zeta)}{\mathcal{Y}_{0,0}(\zeta)}
                +\frac{1}{\zeta}
        \right]            
                \orddif{\mathcal{Y}_{0,0}(\omega)}{J(z)}
        \orddif{E_{\zeta,\bar{\zeta}}}{\omega}
                \frac{1}{\mathcal{Y}_{0,0}(\omega)}\\
&\hs\hs-\frac{1}{4}\oint_{\mathcal{C}_\omega}\frac{\de \zeta}{2\pi\im}
        \oint_{\mathcal{C}}
        \frac{\de \omega}{2\pi\im}
        \left[
                \frac{\mathcal{Y}'_{0,0}(\zeta)}{\mathcal{Y}_{0,0}(\zeta)}
                +\frac{1}{\zeta}
        \right]
                \orddif{E_{\zeta,\bar{\zeta}}}{\omega}
                \frac{1}{\mathcal{Y}_{0,0}(\omega)}
                \orddif{\mathcal{Y}_{0,0}(\omega)}{J(z)}\\
&=\frac{1}{4}\mathop{\oint\oint}_{\mathcal{C}^{(\omega)}>\mathcal{C}^{(\zeta)}}
        \frac{\de \zeta}{2\pi\im}\frac{\de \omega}{2\pi\im}
        \left[
                \frac{\mathcal{Y}'_{0,0}(\zeta)}{\mathcal{Y}_{0,0}(\zeta)}
                +\frac{1}{\zeta}
        \right]
                \orddif{E_{\omega,\bar{\omega}}}{z}     
                \pardif{}{\omega}
        \left[
                \orddif{E_{\zeta,\bar{\zeta}}}{\omega}
                \frac{1}{2\mathcal{Y}_{0,0}(\omega)}
        \right]\\
&\hs\hs-\frac{1}{4}\oint_{\mathcal{C}_\omega}\frac{\de \zeta}{2\pi\im}
        \oint_{\mathcal{C}}
        \frac{\de \omega}{2\pi\im}
        \left[
                \frac{\mathcal{Y}'_{0,0}(\zeta)}{\mathcal{Y}_{0,0}(\zeta)}
                +\frac{1}{\zeta}
        \right]
                \frac{1}{\omega-\zeta}
                \frac{1}{\mathcal{Y}_{0,0}(\omega)}
                \orddif{\mathcal{Y}_{0,0}(\omega)}{J(z)}\\
&=\frac{1}{4}\mathop{\oint\oint}_{\mathcal{C}^{(\omega)}>\mathcal{C}^{(\zeta)}}
        \frac{\de \zeta}{2\pi\im}\frac{\de \omega}{2\pi\im}
                \orddif{E_{\omega,\bar{\omega}}}{z}     
                \left[
                \pardif{}{\omega}-\frac{\mathcal{Y}'_{0,0}(\omega)}{\mathcal{Y}_{0,0}(\omega)}
        \right]
        \left[
                \orddif{E_{\zeta,\bar{\zeta}}}{\omega}
                \frac{1}{2\mathcal{Y}_{0,0}(\omega)}
        \right]
        \left[
                \frac{\mathcal{Y}'_{0,0}(\zeta)}{\mathcal{Y}_{0,0}(\zeta)}
                +\frac{1}{\zeta}
        \right]\\
&\hs\hs+\frac{1}{4}
        \oint_{\mathcal{C}}
        \frac{\de \omega}{2\pi\im}
        \left[
                \frac{\mathcal{Y}'_{0,0}( \omega)}{\mathcal{Y}_{0,0}( \omega)}
                +\frac{1}{ \omega}
        \right]  
                \frac{1}{\mathcal{Y}_{0,0}(\omega)}
                \orddif{\mathcal{Y}_{0,0}(\omega)}{J(z)}\\
&=w_{0,2}^{(a)}(z)-\frac{1}{4}
        \oint_{\mathcal{C}}
        \frac{\de \omega}{2\pi\im}
        \frac{1}{2}
        \pardif{}{\omega}
        \left[
                \frac{\mathcal{Y}'_{0,0}( \omega)}{\mathcal{Y}_{0,0}( \omega)^2}
                +\frac{1}{ \mathcal{Y}_{0,0}(\omega)\omega}
        \right]  
                \orddif{E_{\omega,\bar{\omega}}}{z}\\
&=w_{0,2}^{(a)}(z)-\frac{1}{4}
        \oint_{\mathcal{C}}
        \frac{\de \omega}{2\pi\im}
        \frac{1}{2}
        \left\{
        \left[
                \frac{\mathcal{Y}'_{0,0}( \omega)}{\mathcal{Y}_{0,0}( \omega)^2}
        \right]'
        -\frac{\mathcal{Y}'_{0,0}( \omega)}{ \mathcal{Y}_{0,0}(\omega)^{2}\omega}
        -\frac{1}{ \mathcal{Y}_{0,0}(\omega)\omega^2}
        \right\}  
                \orddif{E_{\omega,\bar{\omega}}}{z}.
\end{split}
\end{align}

\subsubsection{$w_{0,2}^{(c)}$}
By using eq.(\ref{eq:dE/dj_vs_res}),
        we obtain
\begin{align}
\begin{split}
w_{0,2}^{(c)}(z)
&=\frac{1}{4}\oint_\mathcal{C}
        \frac{\de \zeta}{2 \pi \im}
        \left[
                \frac{\mathcal{Y}'_{0,0}(\zeta)}{\mathcal{Y}_{0,0}(\zeta)}
                +\frac{1}{\zeta}
        \right]
    \int'_\mathcal{C}
        \frac{\de \omega}{2 \pi \im}
        \log \left[
                \omega\mathcal{Y}_{0,0}(\omega)
        \right]\\
&\hs\hs\times\sum_{i=1}^4\Res_{\eta\to x_i}
        \frac{1}{2\mathcal{Y}_{0,0}(\eta)}
        \orddif{E_{\eta,\bar{\eta}}}{z}
        \orddif{E_{\zeta,\bar{\zeta}}}{\eta}
        \left[
                B(\omega,\eta)-B(\omega,\bar{\eta})
        \right]\\
&=-\frac{1}{4}\mathop{\oint\oint\oint}_{\mathcal{C}^{(\zeta)}>\mathcal{C}^{(\omega)}>\mathcal{C}^{(\eta)}}
        \frac{\de \zeta}{2 \pi \im}
        \frac{\de \omega}{2 \pi \im}
        \frac{\de \eta}{2 \pi \im}
    \left[
        \frac{
                \mathcal{Y}'_{0,0}(\zeta)}{\mathcal{Y}_{0,0}(\zeta)
        }
        +\frac{1}{\zeta}
    \right]
    \left[
        \frac{
                \mathcal{Y}'_{0,0}(\omega)}{\mathcal{Y}_{0,0}(\omega)
        }
        +\frac{1}{\omega}
    \right]\\
&\hs\hs \times
        \frac{1}{2\mathcal{Y}_{0,0}(\eta)}
        \orddif{E_{\eta,\bar{\eta}}}{z}
        \orddif{E_{\zeta,\bar{\zeta}}}{\eta}
        \orddif{E_{\omega,\bar{\omega}}}{\eta}.
\end{split}
\end{align}
Let $\mathfrak{w}(\zeta,\omega;\eta)$ be
\begin{align}
\begin{split}
\mathfrak{w}(\zeta,\omega;\eta)&\equiv\mathfrak{w}(\omega,\zeta;\eta)\\
&=- \frac{1}{4(2 \pi \im)^3}
    \left[
        \frac{
                \mathcal{Y}'_{0,0}(\zeta)}{\mathcal{Y}_{0,0}(\zeta)
        }
        +\frac{1}{\zeta}
    \right]
    \left[
        \frac{
                \mathcal{Y}'_{0,0}(\omega)}{\mathcal{Y}_{0,0}(\omega)
        }
        +\frac{1}{\omega}
    \right]
        \frac{1}{2\mathcal{Y}_{0,0}(\eta)}
        \orddif{E_{\eta,\bar{\eta}}}{z}
        \orddif{E_{\zeta,\bar{\zeta}}}{\eta}
        \orddif{E_{\omega,\bar{\omega}}}{\eta}.
\end{split}
\end{align}
By using $\mathfrak{w}(\zeta,\omega;\eta)$,
    $w_{0,2}^{(c)}$ is written as
\begin{align}
\begin{split}
w_{0,2}^{(c)}
&=\mathop{\oint\oint\oint}_{
        \mathcal{C}^{(\zeta)}>\mathcal{C}^{(\omega)}>\mathcal{C}^{(\eta)}
        }
        \mathfrak{w}(\zeta,\omega;\eta)\de\zeta\de\omega\de\eta\\
&=\left[
        \mathop{\oint\oint\oint}_{
                \mathcal{C}^{(\zeta)}>\mathcal{C}^{(\eta)}>\mathcal{C}^{(\omega)}
        }\de\zeta\de\omega\de\eta
        +\mathop{\oint\oint}_{
                \mathcal{C}^{(\zeta)}>\mathcal{C}^{(\eta)}
        }
        \de\zeta\de\eta
        \oint_{\mathcal{C}_\eta}
        \de\omega
    \right]
    \mathfrak{w}(\zeta,\omega;\eta)\\
&=\left[
        \mathop{\oint\oint\oint}_{
                \mathcal{C}^{(\eta)}>\mathcal{C}^{(\zeta)}>\mathcal{C}^{(\omega)}
        }\de\eta\de\zeta\de\omega
        +\mathop{\oint\oint}_{
                \mathcal{C}^{(\eta)}>\mathcal{C}^{(\omega)}
        }
        \de\eta\de\omega
        \oint_{\mathcal{C}_\eta}
        \de\zeta
        +\mathop{\oint\oint}_{
                \mathcal{C}^{(\zeta)}>\mathcal{C}^{(\eta)}
        }
        \de\zeta\de\eta
        \oint_{\mathcal{C}_\eta}
        \de\omega
    \right]
    \mathfrak{w}(\zeta,\omega;\eta)\\
&=\Bigg[
        \mathop{\oint\oint\oint}_{
                \mathcal{C}^{(\eta)}>\mathcal{C}^{(\zeta)}>\mathcal{C}^{(\omega)}
        }\de\eta\de\zeta\de\omega
        +\mathop{\oint\oint}_{
                \mathcal{C}^{(\eta)}>\mathcal{C}^{(\omega)}
        }
        \de\eta\de\omega
        \oint_{\mathcal{C}_\eta}
        \de\zeta
        +\mathop{\oint\oint}_{
                \mathcal{C}^{(\eta)}>\mathcal{C}^{(\zeta)}
        }
        \de\eta\de\zeta
        \oint_{\mathcal{C}_\eta}
        \de\omega\\
&\hs\hs\hs\hs\hs\hs\hs\hs\hs+\oint_{\mathcal{C}}\de\eta
                \oint_{
                        \mathcal{C}_{\eta}
                }\de\zeta
                \oint_{
                        \mathcal{C}_{\eta}
                }\de\omega\Bigg]
    \mathfrak{w}(\zeta,\omega;\eta).
\end{split}
\end{align}
The first, second, third, and the fourth term of the last equation
    is respectively given by
\begin{align}
\begin{split}
\mathrm{The \  first \  term}=-\oint_{\mathcal{C}}
        \frac{\de \eta}{2 \pi \im}      
        \frac{1}{2\mathcal{Y}_{0,0}(\eta)}
        \orddif{E_{\eta,\bar{\eta}}}{z}
            w_{0,1}(\eta)^2,
\end{split}
\end{align}
\begin{align}
\begin{split}
\mathrm{The \  second \  term}&=\mathrm{The \  third \  term}\\
&=-\frac{1}{4}\mathop{\oint\oint}_{
                \mathcal{C}^{(\eta)}>\mathcal{C}^{(\zeta)}
        }
        \frac{\de\eta}{2 \pi \im}
        \frac{\de\zeta}{2 \pi \im}
        \oint_{\mathcal{C}_\eta}     
        \frac{\de\omega}{2 \pi \im}
    \left[
        \frac{
                \mathcal{Y}'_{0,0}(\zeta)}{\mathcal{Y}_{0,0}(\zeta)
        }
        +\frac{1}{\zeta}
    \right]
    \left[
        \frac{
                \mathcal{Y}'_{0,0}(\omega)}{\mathcal{Y}_{0,0}(\omega)
        }
        +\frac{1}{\omega}
    \right]\\
&\hs\hs\times        \frac{1}{2\mathcal{Y}_{0,0}(\eta)}
        \orddif{E_{\eta,\bar{\eta}}}{z}
        \orddif{E_{\zeta,\bar{\zeta}}}{\eta}
        \frac{1}{\eta-\omega}\\
&=\frac{1}{4}\mathop{\oint\oint}_{
                \mathcal{C}^{(\eta)}>\mathcal{C}^{(\zeta)}
        }
        \frac{\de\eta}{2 \pi \im}
        \frac{\de\zeta}{2 \pi \im}
    \left[
        \frac{
                \mathcal{Y}'_{0,0}(\zeta)}{\mathcal{Y}_{0,0}(\zeta)
        }
        +\frac{1}{\zeta}
    \right]
    \left[
        \frac{
                \mathcal{Y}'_{0,0}(\eta)}{\mathcal{Y}_{0,0}(\eta)
        }
        +\frac{1}{\eta}
    \right]
    \frac{1}{2\mathcal{Y}_{0,0}(\eta)}
        \orddif{E_{\eta,\bar{\eta}}}{z}
        \orddif{E_{\zeta,\bar{\zeta}}}{\eta}\\
&=-\frac{1}{2}\mathop{\oint\oint}_{
                \mathcal{C}^{(\eta)}>\mathcal{C}^{(\zeta)}
        }
        \frac{\de\eta}{2 \pi \im}
    \frac{1}{2\mathcal{Y}_{0,0}(\eta)}
    \orddif{E_{\eta,\bar{\eta}}}{z}
    \left[
        \frac{
                \mathcal{Y}'_{0,0}(\eta)}{\mathcal{Y}_{0,0}(\eta)
        }
        +\frac{1}{\eta}
    \right]        
        w_{0,1}(\eta),
\end{split}
\end{align}
\begin{align}
\begin{split}
\mathrm{The \ fourth \  term}
&=-\frac{1}{4}\oint_{\mathcal{C}}\frac{\de\eta}{2 \pi \im}
                \oint_{
                        \mathcal{C}_{\eta}
                }\frac{\de\zeta}{2 \pi \im}
                \oint_{
                        \mathcal{C}_{\eta}
                }\frac{\de\omega}{2 \pi \im}    
    \left[
        \frac{
                \mathcal{Y}'_{0,0}(\zeta)}{\mathcal{Y}_{0,0}(\zeta)
        }
        +\frac{1}{\zeta}
    \right]
    \left[
        \frac{
                \mathcal{Y}'_{0,0}(\omega)}{\mathcal{Y}_{0,0}(\omega)
        }
        +\frac{1}{\omega}
    \right]\\
&\hs\hs\hs    \times
        \frac{1}{2\mathcal{Y}_{0,0}(\eta)}
        \orddif{E_{\eta,\bar{\eta}}}{z}
        \frac{1}{\eta-\zeta}
        \frac{1}{\eta-\omega}\\
&=-\frac{1}{4}\oint_\mathcal{C}
        \frac{\de\eta}{2 \pi \im}
    \left[
        \frac{
                \mathcal{Y}'_{0,0}(\eta)}{\mathcal{Y}_{0,0}(\eta)
        }
        +\frac{1}{\eta}
    \right]
    \left[
        \frac{
                \mathcal{Y}'_{0,0}(\eta)}{\mathcal{Y}_{0,0}(\eta)
        }
        +\frac{1}{\eta}
    \right]
        \frac{1}{2\mathcal{Y}_{0,0}(\eta)}
        \orddif{E_{\eta,\bar{\eta}}}{z}.
\end{split}
\end{align}
\subsubsection{The sum}
The sum of $w_{0,2}^{(a)}(z)$,
    $w_{0,2}^{(b)}(z)$, and
    $w_{0,2}^{(c)}(z)$
    is given by 
\begin{align}
\begin{split}
&w_{0,2}^{(a)}(z)+w_{0,2}^{(b)}(z)+w_{0,2}^{(c)}(z)\\
&=\oint_{\mathcal{C}}\frac{\de \omega}{2\pi\im}
        \frac{1}{2\mathcal{Y}_{0,0}(\omega)}
        \orddif{
                E_{\omega,\bar{\omega}}
        }{z}\\
&\hs\times
        \Bigg\{
                -\left[
                        \pardif{}{\omega}
                        -\frac{\mathcal{Y'}_{0,0}(\omega)}{\mathcal{Y}_{0,0}(\omega)}
                \right]w_{0,1}(\omega)
-w_{0,1}(\omega)^2
-\left[
                        \frac{\mathcal{Y'}_{0,0}(\omega)}{\mathcal{Y}_{0,0}(\omega)}
                        +\frac{1}{\omega}
                \right]w_{0,1}(\omega)\\
&\hs-\frac{1}{4}\left[
                        \frac{\mathcal{Y''}_{0,0}(\omega)}{\mathcal{Y}_{0,0}(\omega)}
                        -\frac{2\mathcal{Y}'_{0,0}(\omega)^{2}}{\mathcal{Y}_{0,0}(\omega)^{2}}
                        -\frac{\mathcal{Y}'_{0,0}(\omega)}{\omega\mathcal{Y}_{0,0}(\omega)}-\frac{1}{\omega^{2}}
                +\frac{\mathcal{Y}'_{0,0}(\omega)^{2}}{\mathcal{Y}_{0,0}(\omega)^{2}}+\frac{2\mathcal{Y}'_{0,0}(\omega)^{}}{\omega\mathcal{Y}_{0,0}(\omega)^{}}+\frac{1}{\omega^{2}}
                \right]
        \Bigg\}\\
&=\oint_{\mathcal{C}}\frac{\de \omega}{2\pi\im}
        \frac{1}{2\mathcal{Y}_{0,0}(\omega)}
        \orddif{
                E_{\omega,\bar{\omega}}
        }{z}
        \left\{2\mathcal{Y}_{0,0}(\omega)w_{0,2}(\omega)
        -\frac{1}{4}\left[\frac{\mathcal{Y}'_{0,0}(\omega)}{\mathcal{Y}_{0,0}(\omega)}\right]
        '-\frac{1}{4\omega}\frac{\mathcal{Y}'_{0,0}(\omega)}{\mathcal{Y}_{0,0}(\omega)}
        \right\}\\
&=w_{0,2}(\omega)+\frac{1}{8}\oint_{\mathcal{C}}\frac{\de \omega}{2\pi\im}
        \frac{1}{2\mathcal{Y}_{0,0}(\omega)}
        \orddif{
                E_{\omega,\bar{\omega}}
        }{z}
        \sum_{i=1}^4\left\{\frac{1}{(\omega-x_i)^2}
        -\frac{1}{\omega(\omega-x_i)}
        \right\}\\
&=w_{0,2}(\omega)+\frac{1}{8}\oint_{\mathcal{C}}\frac{\de \omega}{2\pi\im}
        \frac{1}{2\mathcal{Y}_{0,0}(\omega)}
        \orddif{
                E_{\omega,\bar{\omega}}
        }{z}
        \sum_{i=1}^4\left\{\frac{1}{(\omega-x_i)^2}
        -\frac{1}{x_i(\omega-x_i)}
        \right\},
\end{split}
\end{align}
where we have used eqs.(\ref{eq:loop_eq_of_lower_ij}), (\ref{eq:int_E_w=w}), and (\ref{eq:int_E/2y_f=0}).

\subsection{Akemann's formula}
Let $\chi^{(1)}_i(z)$ and $\chi^{(1)}_i(z)$ be
\begin{align}
\begin{split}
\chi^{(1)}_i(z)&=\orddif{x_i}{J(z)},\\
\chi^{(2)}_i(z)
&=-\frac{1}{3}
        \orddif{\ \ \ }{J(z)} 
        \left[
                2\log \abs{M_i}+\sum_{ j\neq i} \log \abs{x_i-x_j}
        \right].
\end{split}
\end{align}
Akemann\cite{Akemann:1996zr} has shown that
        $\chi^{(k)}_i(z)$ satisfies
\begin{align}
\begin{split}
\oint_{\mathcal{C}}\frac{\de\omega}{2\pi\im}\frac{W'_{0,0}(\omega)}{\omega-z}
        \left[
                \chi^{(k)}_i(\omega)-
                \sqrt{\prod_{i=1}^4\frac{z-x_i}{\omega-x_i}}\chi^{(k)}_i(z)
        \right]&=\frac{1}{(z-x_i)^{k}}.
\end{split}
\end{align}
Chekhov and Eynard have shown\cite{CE0504,CE0604}
\begin{align}
\begin{split}
\chi^{(k)}_i(z)=-\mathop{\oint\oint}_{\mathcal{C}^{(\zeta)}>\mathcal{C}^{(\omega)}}\frac{\de\zeta}{2\pi\im}\frac{\de\omega}{2\pi\im}
        \frac{1}{2\mathcal{Y}_{0,0}(\zeta)}
        \orddif{E_{\zeta,\bar{\zeta}}}{z}
        \frac{W'_{0,0}(\omega)}{\omega-\zeta}
        \left[
                \chi^{(k)}_i(\omega)-
                \sqrt{\prod_{i=1}^4\frac{\zeta-x_i}{\omega-x_i}}\chi^{(k)}_i(\zeta)
        \right],
\end{split}
\end{align}
which implies
\begin{align}
\begin{split}
\chi^{(k)}_i(z)=-\oint_{\mathcal{C}}\frac{\de\zeta}{2\pi\im}
        \frac{1}{2\mathcal{Y}_{0,0}(\zeta)}
        \orddif{E_{\zeta,\bar{\zeta}}}{z}
        \frac{1}{(\zeta-x_i)^k}.
\end{split}
\end{align}
We thus obtain
\begin{align}
\begin{split}
\orddif{F_{0,2}^{(1)}}{J(z)}
&=-\frac{1}{8}
        \sum_{i=1}^4 
                \frac{1}{x_i}\orddif{x_i}{J(z)}
=-\frac{1}{8}
        \sum_{i=1}^4 
                \frac{\chi^{(1)}_i(z)}{x_i}\\
&=\frac{1}{8}\oint_{\mathcal{C}}\frac{\de\zeta}{2\pi\im}
        \frac{1}{2\mathcal{Y}_{0,0}(\zeta)}
        \orddif{E_{\zeta,\bar{\zeta}}}{z}
        \sum_{i=1}^4\frac{1}{x_i(\zeta-x_i)},\\
\orddif{F_{0,2}^{(2)}}{J(z)}
&=-\frac{1}{12}
        \sum_{i=1}^4 
        \left[
                \log \abs{M_i}+\sum_{1\le j <i \le 4} \log \abs{x_i-x_j}
        \right]=\frac{1}{8}
        \sum_{i=1}^4 \chi^{(2)}_i(z)\\
&=-\frac{1}{8}\oint_{\mathcal{C}}\frac{\de\zeta}{2\pi\im}
        \frac{1}{2\mathcal{Y}_{0,0}(\zeta)}
        \orddif{E_{\zeta,\bar{\zeta}}}{z}
        \sum_{i=1}^4\frac{1}{(\zeta-x_i)^{2}}.
\end{split}
\end{align}
We therefore obtain $\orddif{\ \ \ }{J(z)}F_{0,2}\Big|_{J=0} = w_{0,2}(z)$.

\section{Expression for $w_{0,0}(z)$}
\label{sec:Appendix_On_Planar_limit}
In this appendix, we study the planar limit.
The lowest order of the $z$-expansion of Virasoro constraints (\ref{eq:loop_equation_0}) implies  
\begin{align}
\begin{split}
0&= \frac{1}{Z} \mathop{\int}_{\mathcal{D}_L\times\mathcal{D}_R} 
        \prod_{I'=1}^N \de \lambda_{I'}\  
        \sum_{I=1}^N\pardif{}{\lambda_I  }
    \left(
        \Delta(\lambda_J)^{2b_E^2}  
        \exp\left[
                b_E\sum_{\bar{I}=1}^N W(\lambda_{\bar{I}};\alpha_i,\mathrm{g}_\ell)
        \right]
    \right)\\
&=
    (\bar{\alpha}_1 +g_\ell Q_E)w(0)
    +\bar{\alpha}_2 w(q)
    +\bar{\alpha}_3 w(1)
.
\end{split}
\end{align}
We thus obtain
\begin{align}
\begin{split}
0&=\bar{\alpha}_1 w_{0,0}(0)
    +\bar{\alpha}_2 w_{0,0}(q)
    +\bar{\alpha}_3 w_{0,0}(1),\\
0&=\bar{\alpha}_1 w_{i,j}(0)
    +\bar{\alpha}_2 w_{i,j}(q)
    +\bar{\alpha}_3 w_{i,j}(1)+w_{i,j-1}(0)
    .
\end{split}
\label{eq:leading_of_virasoro_c}
\end{align}
Using eq.(\ref{eq:leading_of_virasoro_c}),
we can rewrite $f_{0,0}(z)$ as
\begin{align}
    \begin{split}
    f_{0,0}(z) &=
    \frac{
        q \bar{\alpha}_1 w_{0,0}(0)
       +z \left[
           q\bar{\alpha}_2 w_{0,0}(q)
           + \bar{\alpha}_3 w_{0,0}(1)
       \right]
    }{z(z-q)(z-1)}
\end{split}.
\end{align}
Eq.(\ref{eq:planar result}) can be written as
\begin{align}
w_{0,0}(z) &= -\frac{1}{2}\Big\{ W'_{0,0}(z) - \sqrt{[W'_{0,0}(z)]^2+4f_{0,0}(z)} \Big\}\\
\begin{split}
&=-\frac{1}{2}\Bigg\{
        \frac{\bar{\alpha}_1}{z} + \frac{\bar{\alpha}_2}{z-q} + \frac{\bar{\alpha}_3}{z-1}\\
&\hs -\sqrt{
            \Big[
                \frac{\bar{\alpha}_1}{z}
                + \frac{\bar{\alpha}_2}{z-q}
                + \frac{\bar{\alpha}_3}{z-1}
            \Big]^2
            + 4 \frac{
                q \bar{\alpha}_1 w_{0,0}(0)
                +z \left[
                        q\bar{\alpha}_2 w_{0,0}(q)
                        + \bar{\alpha}_3 w_{0,0}(1)
                \right]
            }{z(z-q)(z-1)} 
        }
    \Bigg\}.
\end{split}
\label{eq:w00_with_w00(0,q,1)}
\end{align}
Taking the limit of (\ref{eq:w00_with_w00(0,q,1)}) as $z\to\infty$, we obtain
\begin{align}
\begin{split}
\varrho&=\lim_{z\to\infty} z w_{0}(z)\\
&=-\frac{1}{2}\Bigg\{
        \bar{\alpha}_1 + \bar{\alpha}_2 + \bar{\alpha}_3
        -\sqrt{
            \Big(
                \bar{\alpha}_1 
                +\bar{\alpha}_2
                +\bar{\alpha}_3
            \Big)^2
            + 4 \left[
                q\bar{\alpha}_2 w_{0,0}(q)
                + \bar{\alpha}_3 w_{0,0}(1)
            \right]
        }
    \Bigg\}.
\end{split}
\end{align}
where we have used $w_{0,0}(z) \to \frac{\varrho}{z} \ \ (z\to\infty)$.
The above equation reduces to
\begin{align}
q \bar{\alpha}_2  w_{0,0}(q)+\bar{\alpha}_3 w_{0,0}(1)=-(m_2+m_4)(m_2+m_3).
\end{align}
By using the above equation, $f_{0,0}(z)$ can be written as
\begin{align}
f_{0,0}(z)=\frac{q (m_2-m_1)w_{0,0}(0)-z (m_2+m_4)(m_2+m_3)
}{z(z-q)(z-1)}.
\end{align}

So far, we have expressed $w_{0,0}$
    in terms of $\bar{\alpha}_i$ and $m_i$.
Alternatively, we may express the root in terms of the cut variables $x_i$
Comparing these two, we will obtain a set of relations satisfied by $x_i$
    as we will see below shortly.
After factorizing the root of $w_{0,0}(z)$, we obtain
\begin{align}
\begin{split}
w_{0,0}(z)
&=-\frac{1}{2}\Bigg\{
        \frac{\bar{\alpha}_1}{z} + \frac{\bar{\alpha}_2}{z-q} + \frac{\bar{\alpha}_3}{z-1}
            -\frac{
                \bar{\alpha}_4
            }{
                z(z-1)(z-q)
            }
        \sqrt{
            \prod_{i=1}^4(z-x_i)
        }
    \Bigg\},
\end{split}\label{eq:w_00_wrt_a_x}
\end{align}
where
\begin{align}
x_1<x_2<x_3<x_4.
\end{align}
By using eq. (\ref{eq:w_00_wrt_a_x}),
    we obtain
\begin{align}
\begin{split}
\sqrt{
    \prod_{i=1}^4€x_i
}&=
\frac{q \bar{\alpha} _1}{\bar{\alpha}_4},\\
\sqrt{
    \prod_{i=1}^4(x_i-q)
}&=
-\frac{(1-q) q \bar{\alpha} _2}{\bar{\alpha}_4},\\
\sqrt{
    \prod_{i=1}^4(x_i-1)
}&=
\frac{(1-q) \bar{\alpha} _3}{\bar{\alpha}_4}.
\end{split}\label{eq:pre_q_y_relation}
\end{align}
Eq.(\ref{eq:pre_q_y_relation}) has been shown in \cite{SW0911}. 
By using eq.(\ref{eq:pre_q_y_relation}),
        we have
\begin{align}
\begin{split}
w_{0,0}(z)
&=-\frac{\bar{\alpha}_4}{2}\Bigg\{
        \lim_{\zeta\to 0}\frac{
                \sqrt{
            \prod_{i=1}^4(\zeta-x_i)
        }
            }{
                z(\zeta-q)(\zeta-1)
            }
        +
        \lim_{\zeta\to q}\frac{
                \sqrt{
            \prod_{i=1}^4(\zeta-x_i)
        }
            }{
                \zeta(z-q)(\zeta-1)
            }
        \\
        &\hs\hs\hs\hs+ \lim_{\zeta\to1}\frac{
                \sqrt{
            \prod_{i=1}^4(\zeta-x_i)
        }
            }{
                \zeta(\zeta-q)(z-1)
            }   
            -\frac{
                \sqrt{
            \prod_{i=1}^4(z-x_i)
        }
            }{
                z(z-q)(z-1)
            }    
    \Bigg\}.
\end{split}\label{eq:w_00_wrt_x_q}
\end{align}
Eq.(\ref{eq:w_00_wrt_x_q}) implies
\begin{align}
\begin{split}
w_{0,0}(0)
&=\frac{\bar{\alpha}_1}{2}\left(
                1+\frac{1}{q}-\frac{1}{2}\sum_{i=1}^4 \frac{1}{x_i}
        \right)
        +\frac{\bar{\alpha}_2}{2q}+\frac{\bar{\alpha}_3}{2},\\
w_{0,0}(q)
&=\frac{\bar{\alpha}_2}{2}\left(
                \frac{1}{1-q}-\frac{1}{q}-\frac{1}{2}\sum_{i=1}^4 \frac{1}{x_i-q}
        \right)
        -\frac{\bar{\alpha}_1}{2q}+\frac{\bar{\alpha}_3}{2(1-q)},\\
w_{0,0}(1)
&=\frac{\bar{\alpha}_3}{2}\left(
                -1-\frac{1}{1-q}-\frac{1}{2}\sum_{i=1}^4 \frac{1}{x_i-1}
        \right)
        -\frac{\bar{\alpha}_1}{2}-\frac{\bar{\alpha}_2}{2(1-q)}.
\end{split}
\end{align} 
By using (\ref{eq:DV_prescription_with_expansion}), we have
\begin{align}
\begin{split}
\frac{\bar{\alpha}_4}{2}\oint_{\mathcal{C}_k} \frac{\de \omega}{2\pi\im}\frac{             
        \sqrt{
            \prod_{i=1}^4( \omega-x_i)
        }
            }{
                 \omega( \omega-1)( \omega-q)
            }
=
\int_{x_{2k-1}}^{x_{2k}} \frac{             
        \sqrt{
            \abs{\prod_{i=1}^4( \omega-x_i)}
        }
            }{
                 \abs{\omega( \omega-1)( \omega-q)}
            }\frac{\bar{\alpha}_4\de \omega}{2\pi}
&=\varrho\left[
        \frac{N_L}{N}\delta_{k,1}+\frac{N_R}{N}\delta_{k,2}
        \right].
\end{split}
\label{eq:boundary_c_from_DV_prescription}
\end{align}
We conclude that, given $\bar{\alpha}_i$, $\varrho$, $q$, and $N_R/N$,
   it is possible to obtain the seven unknown quantities $w_{0,0}(0)$, $w_{0,0}(q)$, $w_{0,0}(1)$, $x_i$,
   using (D.11), (D.12), (D.14).
While it is hard to solve these, the quantities
   $\prod_{i=1}^4 x_i$ and $\prod_{i=1}^4 M_i$ are relatively easy to obtain.
The former is given in (\ref{eq:pre_q_y_relation})
   while the latter, after
\begin{align}
\begin{split}
M_j&=\oint_{\mathcal{C}}\frac{\de\omega}{2\pi\im}
    \frac{W'_{0,0}(\omega)}{\sqrt{\prod_{i=1}^4(\omega-x_i)}}\frac{1}{\omega-x_j}\\
&=-\oint_{\mathcal{C}_0+\mathcal{C}_q+\mathcal{C}_1}\frac{\de\omega}{2\pi\im}
    \frac{W'_{0,0}(\omega)}{\sqrt{\prod_{i=1}^4(\omega-x_i)}}\frac{1}{\omega-x_j}\\
&=\frac{\bar{\alpha}_1}{x_j\sqrt{\prod_{i=1}^4x_i}}
        +\frac{\bar{\alpha}_2}{(x_j-q)\sqrt{\prod_{i=1}^4(x_i-q)}}
        +\frac{\bar{\alpha}_3}{(x_j-1)\sqrt{\prod_{i=1}^4(x_i-1)}}\\
&=\bar{\alpha}_4 \left[
        \frac{1}{q x_j}-\frac{1}{q (1-q)(x_j-q)}+\frac{1}{(1-q)(x_j-1)}
        \right]\\
&=\frac{\bar{\alpha}_4}{x_j(x_j-q)(x_j-1)},
\end{split}
\end{align}
can be handled by (\ref{eq:pre_q_y_relation}).
The concrete forms are
\begin{align}
\begin{split}
\prod_{j=1}^4 M_j &=\prod_{j=1}^4 \frac{\bar{\alpha}_4}{x_j(x_j-q)(x_j-1)}\\
&=\frac{(\bar{\alpha}_4)^{10}}{q^4 (1-q)^4 (\bar{\alpha}_1\bar{\alpha}_2\bar{\alpha}_3)^2}\\
\prod_{j=1}^4 x_j &= \left(\frac{q \bar{\alpha}_1}{\bar{\alpha}_4}\right)^2.
\end{split}
\label{eq:M_vs_a4_q}
\end{align}


\begin{thebibliography}{99}

\bibitem{ACKM:1993}
J.~Ambjorn, L.~Chekhov, C.~F.~Kristjansen and Y.~Makeenko,
``Matrix Model Calculations beyond the Spherical Limit,'' Nucl.\ Phys.\ {\bf B 404}, 127-172 (1993); Erratum-ibid. B {\bf 449}, 681 (1995) [arXiv:hep-th/9302014].
\bibitem{Akemann:1996zr}
G.~Akemann,
``Higher genus correlators for the Hermitian matrix model with multiple cuts,''
Nucl.\ Phys.\  {\bf B 482}, 403-430 (1996) [arXiv:hep-th/9606004].



\bibitem{David90}
F.~David,
``Loop equations and nonperturbative effects in two-dimensional quantum gravity,''
Mod. Phys. Lett. {\bf A 5}, 1019-1029 (1990).

\bibitem{MM90}
A.~Mironov and A.~Morozov,
``On the orinin of Virasoro constraints in matrix models: lagrangian approach,''
Phys. Lett. {\bf B 252}, 47-52 (1990).

\bibitem{AM90}
J.~Ambj\o rn and Yu.~Makeenko,
``Properties of loop equations for the Hermitian matrix model and for two-dimensional quantum gravity,''
Mod. Phys. Lett. {\bf A 5}, 1753-1763 (1990).

\bibitem{IM91}
H.~Itoyama and Y.~Matsuo,
``Noncritical Virasoro algebra of the $d<1$ matrix model and the quantized string field,''
Phys. Lett. {\bf B 255}, 202-208 (1991).





\bibitem{FKN1991}
M.~Fukuma, H.~Kawai and R.~Nakayama,
``Continuum Schwinger-Dyson Equations And Universal Structures In Two-Dimensional Quantum Gravity,''
Int. J. Mod. Phys. {\bf A 6}, 1385 (1991).

\bibitem{DVV1991}
R.~Dijkgraaf, E.~Verlinde and H.~Verlinde,
``Loop equations and Virasoro constraints in nonperturbative 2-D quantum gravity,''
Nucl. Phys. {\bf B 348}, 435 (1991).




\bibitem{E0407}
B.~Eynard,
``Topological expansion for the 1-hermitian matrix model correlation functions,''
JHEP \textbf{0411}, 031 (2004) [arXiv:hep-th/0407261].

\bibitem{CE0504}
L.~Chekhov and B.~Eynard,
``Hermitian matrix model free energy: Feynman graph technique for all genera,''
JHEP \textbf{03}, 014 (2006)[arXiv:hep-th/0504116].

\bibitem{CE0604}
L.~Chekhov and B.~Eynard,
``Matrix eigenvalue model: Feynman graph technique for all genera,''
JHEP \textbf{12}, 026 (2006)[arXiv:math-ph/0604014].

\bibitem{C1009}
L.~Chekhov, ``Logarithmic potential beta-ensembles and Feynman graphs,'' [arXiv:1009.5940].


\bibitem{AGT}
  L.~F.~Alday, D.~Gaiotto and Y.~Tachikawa,
  ``Liouville Correlation Functions from Four-dimensional Gauge Theories,''
Lett.\ Math.\ Phys.\  {\bf 91}, 167-197 (2010)
  [arXiv:0906.3219 [hep-th]].

\bibitem{Wyllard}
  N.~Wyllard,
  ``$A_{N-1}$ conformal Toda field theory correlation functions from conformal
  $\mathcal{N}=2$ $SU(N)$ quiver gauge theories,''
  JHEP \textbf{0911}, 002 (2009)
  [arXiv:0907.2189 [hep-th]].


\bibitem{DV}
  R.~Dijkgraaf and C.~Vafa,
  ``Toda Theories, Matrix Models, Topological Strings, and $N=2$ Gauge Systems,''
[arXiv:0909.2453 [hep-th]].

\bibitem{IMO}
  H.~Itoyama, K.~Maruyoshi and T.~Oota,
  ``The Quiver Matrix Model and 2d-4d Conformal Connection,''
Prog. Theor. Phys. \textbf{123}, 957-987 (2010)
[arXiv:0911.4244 [hep-th]].



\bibitem{Seiberg:1994rs}
 N.~Seiberg and E.~Witten,
``Electric - magnetic duality, monopole condensation, and confinement in N=2
supersymmetric Yang-Mills theory,''
Nucl.\ Phys.\   {\bf B  426}, 19 (1994)
[Erratum-ibid.\   {\bf B  430}, 485 (1994)]
[arXiv:hep-th/9407087].

\bibitem{SW9408}
N. Seiberg and E. Witten,
``Monopoles, duality and chiral symmetry breaking
in $N=2$ supersymmetric QCD,''
Nucl. Phys. \textbf{B 431}, 484-550 (1994) 
[arXiv:hep-th/9408099].



\bibitem{Integrability}
A.~Gorsky, I.~Krichever, A.~Marshakov, A.~Mironov and A.~Morozov,
Phys.\ Lett.\  {\bf B  355}, 466 (1995)
[arXiv:hep-th/9505035];
M.~Matone,
Phys. Lett. {\bf B357} (1995) 342-348
[arXiv:hep-th/9506102];
E.~Martinec and N.~Warner,
Nucl.\ Phys.\  {\bf B 459}, 97 (1996)
[arXiv:hep-th/9509161];
T.~Nakatsu and K.~Takasaki,
Mod.\ Phys.\ Lett.\   {\bf A  11}, 157 (1996)
[arXiv:hep-th/9509162];
R.~Donagi and E.~Witten,
Nucl.\ Phys.\   {\bf B  460}, 299 (1996)
[arXiv:hep-th/9510101];
T.~Eguchi and S.~K.~Yang,
Mod.\ Phys.\ Lett.\   {\bf A 11}, 131 (1996)
[arXiv:hep-th/9510183];
H.~Itoyama and A.~Morozov,
Nucl.\ Phys.\   {\bf B  477}, 855 (1996)
[arXiv:hep-th/9511126];
H.~Itoyama and A.~Morozov,
Nucl.\ Phys.\  {\bf B  491}, 529 (1997)
[arXiv:hep-th/9512161];
H.~Itoyama and A.~Morozov,
[arXiv:hep-th/9601168];
G.~Bonelli and M.~Matone, 
Phys. Rev. Lett. {\bf 76} (1996) 4107
[arXiv:hep-th/9602174];
G.~Bonelli and M.~Matone,
Phys. Rev. Lett. {bf 77} (1996) 4712
[arXiv:hep-th/9605090];
G.~Bertoldi and M.~Matone,
Phys. Lett. {\bf B 425} (1998) 104-106
[arXiv:hep-th/9712039].

\bibitem{Dijkgraaf:2002s}
R.~Dijkgraaf and C.~Vafa,
``Matrix models, topological strings, and supersymmetric gauge theories,''
Nucl.\ Phys.\  {\bf B  644}, 3 (2002)
[arXiv:hep-th/0206255];
R.~Dijkgraaf and C.~Vafa,
``On geometry and matrix models,''
Nucl.\ Phys.\  {\bf B  644}, 21 (2002)
[arXiv:hep-th/0207106];
R.~Dijkgraaf and C.~Vafa,
``A Perturbative window into nonperturbative physics,''
[arXiv:hep-th/0208048].

\bibitem{whitham}
H.~Itoyama and A.~Morozov,
Nucl. Phys. {\bf B 657}, 53-78 (2003)
[hep-th/0211245];
H.~Itoyama and A.~Morozov,
Phys. Lett. {\bf B 555}, 287-295 (2003)
[hep-th/0211259];
H.~Itoyama and A.~Morozov,
Prog. Theor. Phys. {\bf 109}, 433-463 (2003)
[hep-th/0212032];
L.~Chekhov, A.~Marshakov, A.~Mironov and D.~Vasiliev,
Phys.\ Lett.\  {\bf B 562}, 323 (2003)
[arXiv:hep-th/0301071];
H.~Itoyama and A.~Morozov,
Int. J. Mod. Phys. {\bf A 18}, 5889-5906 (2003)
[hep-th/0301136];
L.~Chekhov and A.~Mironov,
Phys.\ Lett.\  {\bf B 552}, 293 (2003)
[arXiv:hep-th/0209085];
M.~Matone,
Nucl.\ Phys.\  {\bf B 656}, 78 (2003)
[arXiv:hep-th/0212253];
A.~Dymarsky and V.~Pestun,
of the superpotential,''
Phys.\ Rev.\  {\bf D 67}, 125001 (2003)
[arXiv:hep-th/0301135];
S.~Aoyama and T.~Masuda,
JHEP {\bf 0403}, 072 (2004)
[arXiv:hep-th/0309232];
H.~Itoyama and H.~Kanno,
Nucl.\ Phys.\ {\bf B 686}, 155 (2004)
[arXiv:hep-th/0312306].


\bibitem{DF}
  V.~S.~Dotsenko and V.~A.~Fateev,
  ``Conformal algebra and multipoint correlation functions in  2D statistical
  models,''
  Nucl.\ Phys.\  {\bf B 240}, 312-348 (1984); \\
V.~S.~Dotsenko and V.~A.~Fateev,
``Four Point Correlation Functions And The Operator Algebra In The
Two-Dimensional Conformal Invariant Theories With The Central Charge $c \leq 1$,''
Nucl.\ Phys.\  {\bf B 251}, 691-734 (1985).


\bibitem{MMS0911}
  A.~Mironov, A.~Morozov and Sh.~Shakirov,
  ``Matrix Model Conjecture for Exact BS Periods and Nekrasov Functions,''
  JHEP {\bf 1002}, 030 (2010)
  [arXiv:0911.5721 [hep-th]].

\bibitem{MMS1001}
A.~Mironov, A.~Morozov and Sh.~Shakirov,
``Conformal blocks as Dotsenko-Fateev Integral Discriminants,''
Int. J. Mod. Phys. {\bf A 25}, 3173-3207 (2010)
[arXiv:1001.0563 [hep-th]].  
  
\bibitem{IO5}
H.~Itoyama and T.~Oota,
``Method of Generating $q$-Expansion Coefficients for Conformal Block and 
$\mathcal{N}=2$ Nekrasov Function by $\beta$-Deformed Matrix Model,''
  Nucl.\ Phys.\  {\bf B 838}, 298-330 (2010)
[arXiv:1003.2929 [hep-th]].

\bibitem{IOY}
H.~Itoyama, T.~Oota and N.~Yonezawa,
``Massive Scaling Limit of beta-Deformed Matrix Model of Selberg Type,'' Phys. Rev. {\bf D 82}, 085031 (2010) [arXiv:1008.1861].


\bibitem{Nekrasov0206}
N.~Nekrasov,
``Seiberg-Witten Prepotential From Instanton Counting,''
[arXiv:hep-th/0206161].



\bibitem{gai0908}
D.~Gaiotto,
``Asymptotically free $\mathcal{N}=2$ theories
and irregular conformal blocks,''
[arXiv:0908.0307 [hep-th]].

\bibitem{MMM0909a}
A.~Marshakov, A.~Mironov and A.~Morozov,
``On non-conformal limit of the AGT relations.,''
Phys. Lett. \textbf{ B 682}, 125-129 (2009)
[arXiv:0909.2052 [hep-th]].

\bibitem{MMM0909b}
A.~Marshakov, A.~Mironov and A.~Morozov,
``Zamolodchikov asymptotic formula and instanton expansion
in $\mathcal{N}=2$ SUSY $N_f = 2 N_c$ QCD,''
JHEP \textbf{0911}, 048 (2009)
[arXiv:0909.3338 [hep-th]].

\bibitem{AM0910}
V.~Alba and And.~Morozov,
``Non-conformal limit of AGT relation from the 1-point torus conformal block,''
JETP Lett. \textbf{90}, 708-712 (2009)
[arXiv:0911.0363 [hep-th]].

\bibitem{EM1}
T.~Eguchi and K.~Maruyoshi,
``Penner Type Matrix Model and Seiberg-Witten Theory,''
JHEP {\bf 1002}, 022 (2010)
[arXiv:0911.4797 [hep-th]].

\bibitem{1003Yanagida}
S.~Yanagida,
``Whittaker vectors of the Virasoro algebra in terms of Jack symmetric polynomial,''
[arXiv:1003.1049 [math.QA]].

\bibitem{EM2}
T.~Eguchi and K.~Maruyoshi,
``Seiberg-Witten theory, matrix model and AGT relation,''
[arXiv:1006.0828 [hep-th]].

\bibitem{FatLit}
  V.~A.~Fateev and A.~V.~Litvinov,
  ``On AGT conjecture,''
  JHEP {\bf 1002}, 014 (2010)
  [arXiv:0912.0504 [hep-th]].

\bibitem{HJS}
L.~Hadasz, Z.~Jask\'{o}lski and P.~Suchanek,
  ``Proving the AGT relation for $N_f = 0,1,2$ antifundamentals,''
  JHEP {\bf 1006}, 046 (2010)
  [arXiv:1004.1841 [hep-th]].



\bibitem{MMM0907}
A.~Marshakov, A.~Mironov and A.~Morozov,
``On Combinatorial Expansions of Conformal Blocks,''
[arXiv:0907.3946 [hep-th]].

\bibitem{MM0908a}
A.~Mironov and A.~Morozov,
``The Power of Nekrasov Functions,''
Phys. Lett. \textbf{B 680}, 188-194 (2009)
[arXiv:0908.2190 [hep-th]].

\bibitem{MM0908b}
A.~Mironov and A.~Morozov,
``On AGT relation in the case of $U(3)$,''
Nucl. Phys. \textbf{B 825}, 1-37 (2010)
[arXiv:0908.2569 [hep-th]].











\bibitem{MMM1003}
  A.~Mironov, Al.~Morozov and And.~Morozov,
  ``Matrix model version of AGT conjecture and generalized Selberg integrals,''
[arXiv:1003.5752 [hep-th]].

\bibitem{CDV1010}
M.~Cheng, R.~Dijkgraaf and C. Vafa,
``Non-Perturbative Topological Strings And Conformal Blocks,''
[arXiv:1010.4573 [hep-th]].

\bibitem{MMS1011a}
A.~Mironov, A.~Morozov and S.~Shakirov,
``Brezin-Gross-Witten model as 'pure gauge' limit of Selberg integrals,'' [arXiv:1011.3481 [hep-th]].

\bibitem{MMS1011b}
A.~Mironov, A.~Morozov and S.~Shakirov,
``Towards a proof of AGT conjecture by methods of matrix models,'' [arXiv:1011.5629 [hep-th]].

\bibitem{AFLT1012}
V.~Alba, V.~Fateev, A.~Litvinov and G.~Tarnopolsky,
``On combinatorial expansion of the conformal blocks arising from AGT conjecture,''
[arXiv:1012.1312 [hep-th]].












\bibitem{sel}
A.~Selberg,
``Bemerkninger om et multipelt integral,''
Norsk Mat.\ Tidsskr.\ \textbf{26}, 71-78 (1944).

\bibitem{mac1}
I.~G.~Macdonald,
``Commuting differential operators and zonal spherical
functions,''
in \textit{Algebraic Groups Utrecht 1986},
Proceedings of a Symposium in Honour of T.~A.~Springer,
Lecture Notes in Math. \textbf{1271}, 189-200, 
ed. by A.~M.~Cohen, W.~H.~Hesselink, W.~L.~J.~van der Kallen and J.~R.~Strooker,
Springer (1987).

\bibitem{kad}
K.~W.~J.~Kadell,
``The Selberg-Jack symmetric functions,''
Adv. Math. \textbf{130}, 33-102 (1997).

\bibitem{kan}
J.~Kaneko,
``Selberg integrals and hypergeometric functions
associated with Jack polynomials,''
SIAM.\ J.\ Math.\ Anal.\ \textbf{24}, 1086-1110 (1993).

\bibitem{I}
S.~Iguri,
``On a Selberg-Schur integral,''
Lett. Math. Phys. \textbf{89}, 141-158 (2009) \\ {}
[arXiv:0810.5552 [math-ph]].













\bibitem{BT0909}
G.~Bonelli and A.~Tanzini,
``Hitchin systems, $\mathcal{N}=2$ gauge theories and W-gravity,''
Phys.\ Lett.\  {\bf B 691}, 111-115 (2010)
[arXiv:0909.4031 [hep-th]].




\bibitem{KPW}
C.~Koz\c{c}az, S.~Pasquetti and N.~Wyllard,
``A \& B model approaches to surface operators and Toda theories,''
[arXiv:1004.2025 [hep-th]].

\bibitem{MS1004}
  A.~Morozov and S.~Shakirov,
  ``The matrix model version of AGT conjecture and CIV-DV prepotential,''
[arXiv:1004.2917 [hep-th]].

\bibitem{AY1004}
  H.~Awata and Y.~Yamada,
  ``Five-dimensional AGT Relation and the Deformed $\beta$-ensemble,''
[arXiv:1004.5122 [hep-th]].

\bibitem{NX1005}
D.~Nanopoulos and D.~Xie,
``Hitchin Equation, Irregular Singularity, and $N=2$ Asymptotical Free
Theories,''
[arXiv:1005.1350 [hep-th]].

\bibitem{Tai1006}
T.~S.~Tai,
``Triality in $SU(2)$ Seiberg-Witten theory and Gauss hypergeometric
  function,''
[arXiv:1006.0471 [hep-th]].

\bibitem{NX1006}
D.~Nanopoulos and D.~Xie,
``$N=2$ Generalized Superconformal Quiver Gauge Theory,''
[arXiv:1006.3486 [hep-th]].

\bibitem{KMS1007}
  S.~Kanno, Y.~Matsuo and S.~Shiba,
  ``Analysis of correlation functions in Toda theory and AGT-W relation for
  $SU(3)$ quiver,''
  [arXiv:1007.0601 [hep-th]].












\bibitem{MY1009}
K.~Maruyoshi and F.~Yagi,
``Seiberg-Witten curve via generalized matrix model,'' JHEP {\bf 1101}, 042 (2011) [arXiv:1009.5553 [hep-th]].

\bibitem{BMS1010}
  A.~Brini, M.~Mari$\tilde{\textrm{n}}$o and S.~Stevan,
``The uses of the refined matrix model recursion,'' [arXiv:1010.1210 [hep-th]].
\bibitem{MMS1010}
A.~Mironov, A.~Morozov and S.~Shakirov,
``On 'Dotsenko-Fateev' representation of the toric conformal blocks,''
J.\ Phys.\  {\bf A  44}, 085401 (2011) [arXiv:1010.1734 [hep-th]].

\bibitem{MMM1011}
A.~Marshakov, A.~Mironov and A.~Morozov,
``On AGT Relations with Surface Operator Insertion and Stationary Limit of
    Beta-Ensembles,'' Teor.\ Mat.\ Fiz.\  {\bf 164}, 3 (2010) [arXiv:1011.4491 [hep-th]].


\bibitem{BMTY1011}
G.~Bonelli, K.~Maruyoshi, A.~Tanzini and F.~Yagi,
``Generalized matrix models and AGT correspondence at all genera,'' [arXiv:1011.5417 [hep-th]].


\bibitem{MMS1012}
A.~Mironov, A.~Morozov and S.~Shakirov,
``A direct proof of AGT conjecture at beta = 1,'' [arXiv:1012.3137 [hep-th]].

\bibitem{M1101}
A.~Marshakov,
``On Gauge Theories as Matrix Models,'' [arXiv:1101.0676 [hep-th]].

\bibitem{BB1102}
A.~Belavin and V.~Belavin,
``AGT conjecture and Integrable structure of Conformal field theory for $c=1$,''
[arXiv:1102.0343 [hep-th]].

\bibitem{MMPS1103}
A.~Mironov, A.~Morozov, A.~Popolitov and Sh.~Shakirov,
``Resolvents and Seiberg-Witten representation for Gaussian beta-ensemble,''
[arXiv:1103.5470 [hep-th]].


\bibitem{Popolitov1001}
A.~Popolitov,
``On relation between Nekrasov Nekrasovfunctions and BS periods in pure SU(N) case,''
[arXiv:1001.1407 [hep-th]].




\bibitem{FHT1003}
M.~Fujita, Y.~Hatsuda and T.~Tai, 
``Genus-one correction to asymptotically free Seiberg-Witten prepotential from Dijkgraaf-Vafa matrix model,''
JHEP {\bf 1003}, 046 (2010)
[arXiv:0912.2988 [hep-th]].

\bibitem{AM0912}
V.~Alba and And.~Morozov,
``Check of AGT Relation for Conformal Blocks on Sphere,''
Nucl. Phys. {\bf B 840}, 441-468 (2010)
[arXiv:0912.2535 [hep-th]].

\bibitem{Giribet1001}
G.~Giribet,
``On AGT description of N=2 SCFT with $N_f = 4$,''
JHEP {\bf 1001}, 097 (2010)
[arXiv:0912.1930 [hep-th]].

\bibitem{AHNT1003}
I.~Antoniadis, S.~Hohenegger, K.~Narain and T.~Taylor,
``Deformed Topological Partition Function and Nekrasov Backgrounds,''
[arXiv:1003.2832 [hep-th]].

\bibitem{Taki1007}
M.~Taki,
``Surface Operator, Bubbling Calabi-Yau and AGT Relation,''
[arXiv:1007.2524 [hep-th]].

\bibitem{R59}
H.~E.~Rauch,
``Weierstrass points, branch points, and moduli of Riemann surfaces,''
Commun. Pure Appl. Math. {\bf12}, 543-560, (1959).

\bibitem{Fay70}
J.~Fay,
``Theta Functions on Riemann Surfaces,''
Lecture Notes in Mathematics, {\bf 352}, Springer. Verlag (1970).

\bibitem{FK92}
H.~M.~Farkas and I.~Kra,
``Riemann Surfaces,'' 2nd ed.,
Graduate Texts in Mathematics, Springer (1992).

\bibitem{Fay92}
J.~Fay,
``Kernel functions, analytic torsion, and moduli spaces,''
Mem. Amer. Math. Soc. {\bf 96} (1992).

\bibitem{KK08}
A.~Kokotov and D.~Korotkin,
``A new hierarchy of integrable systems associated to Hurwitz spaces,''
Phil. Trans. R. Soc. {\bf A 366}, 1055-1088 (2008).

\bibitem{SW0911}
  R.~Schiappa and N.~Wyllard,
``An $A_r$ threesome: Matrix models, $2d$ CFTs and $4d$ $N=2$ gauge theories,''
[arXiv:0911.5337 [hep-th]].

\end{thebibliography}
\end{document}